\documentclass[aps,twocolumn,superscriptaddress,footinbib,longbibliography]{revtex4-1} 

\usepackage{longtable}
\usepackage{morefloats}
\usepackage[dvips]{graphicx}
\usepackage{color}
\usepackage{epsfig,graphicx,amsfonts,amsbsy}
\usepackage{amsmath,amsfonts,amsthm,amssymb}
\usepackage{appendix}
\usepackage{makeidx}
\usepackage{url}
\usepackage{verbatim}
\usepackage[bookmarksnumbered,pdfpagelabels=true,plainpages=false,colorlinks=true,linkcolor=blue,citecolor=red,urlcolor=blue]{hyperref}
\usepackage[rightcaption]{sidecap}
\usepackage{array}
\usepackage{multirow}
\usepackage{tabularx}
\usepackage{braket} 
\usepackage{dsfont}
\usepackage{bbold}
\usepackage{etoolbox}

\usepackage{dcolumn}
\usepackage{bm}
\usepackage{blindtext}
\usepackage{comment}

\AtBeginDocument{%
    \newwrite\bibnotes
    \def\bibnotesext{Notes.bib}
    \immediate\openout\bibnotes=\jobname\bibnotesext
    \immediate\write\bibnotes{@CONTROL{REVTEX41Control}}
    \immediate\write\bibnotes{@CONTROL{%
    apsrev41Control,author="08",editor="1",pages="0",title="0",year="1"}}
     \if@filesw
     \immediate\write\@auxout{\string\citation{apsrev41Control}}%
    \fi
}%

\newcommand{\bs}[1]{\boldsymbol{#1}}
\newcommand{\ssf}[1]{\mathsf{#1}}

\def\Br{{\bs{r}}}
\def\Bk{{\bs{k}}}
\def\BS{{\bs{S}}}
\def\BD{{\bs{D}}}

\def\BM{{\bs{M}}}
\def\BH{{\bs{H}}}

\def\Be{{ \bs{e} }}
\def\Bm{{ \bs{m} }}

\def\xhat{{ \bs{\hat{x}} }}
\def\yhat{{ \bs{\hat{y}} }}
\def\zhat{{ \bs{\hat{z}} }}

\def\SS{{\ssf{S}}}

\def\CalE{{\mathcal{E}}}

\def\Hsw{{H_{\mbox{\tiny SW}}}}

\begin{document}

\title{Magnonic Quadrupole Topological Insulator in Antiskyrmion Crystals}

\author{Tomoki Hirosawa}
\affiliation{Department of Physics, University of Tokyo, Bunkyo, Tokyo 113-0033, Japan}

\author{Sebasti{\'a}n A. D{\'i}az}
\affiliation{Department of Physics, University of Basel, Klingelbergstrasse 82, CH-4056 Basel, Switzerland}

\author {Jelena Klinovaja}
\affiliation{Department of Physics, University of Basel, Klingelbergstrasse 82, CH-4056 Basel, Switzerland}

\author{Daniel Loss}
\affiliation{Department of Physics, University of Basel, Klingelbergstrasse 82, CH-4056 Basel, Switzerland}

\date{\today}


\begin{abstract}
When the crystalline symmetries that protect a higher-order topological phase are not preserved at the boundaries of the sample, gapless hinge modes or in-gap corner states cannot be stabilized. Therefore, careful engineering of the sample termination is required. Similarly, magnetic textures, whose quantum fluctuations determine the supported magnonic excitations, tend to relax to new configurations that may also break crystalline symmetries when boundaries are introduced. Here we uncover that antiskyrmion crystals provide an experimentally accessible platform to realize a magnonic topological quadrupole insulator, whose hallmark signature are robust magnonic corner states. Furthermore, we show that tuning an applied magnetic field can trigger the self-assembly of antiskyrmions carrying a fractional topological charge along the sample edges. Crucially, these fractional antiskyrmions restore the symmetries needed to enforce the emergence of the magnonic corner states. Using the machinery of nested Wilson loops, adapted to magnonic systems supported by noncollinear magnetic textures, we demonstrate the quantization of the bulk quadrupole moment, edge dipole moments, and corner charges.
\end{abstract}

\maketitle


\section{Introduction}

Higher-order topological phases of matter in a $d$-dimensional system are characterized by the presence of in-gap states  that belong to boundaries of dimension lower than $(d-1)$, namely, to hinges or corners \cite{benalcazar_quantized_2017,benalcazar_electric_2017,Schindler2018}. Such states could be protected by crystalline symmetries, sometimes in conjunction with time reversal, or, alternatively, by particle-hole symmetry if superconductivity is involved \cite{volpez_Second_2019, franca_Phase_2019, langbehn_Reflection_2017, wang_Weak_2018, plekhanov_Hinge_2020}.
Although they were initially postulated in electronic systems, they have been extended to include also bosonic excitations such as phonons \cite{serra-garcia_observation_2018,ni_observation_2019,fan_elastic_2019,xue_acoustic_2019} and photons \cite{xie_second-order_2018,chen_direct_2019,xie_visualization_2019,mittal_photonic_2019}. A stringent requirement for the realization of these topological states is the local preservation of the protecting symmetries at the corresponding higher-order boundaries. Therefore, experimental realizations of higher-order topology usually involve careful engineering of the sample boundaries.

Another bosonic excitation in condensed matter are magnons, the quanta of spin waves. While theoretical predictions of first-order topological magnonic states are abundant \cite{shindou_topological_2013, Zhang2013, mook_Edge_2014, chisnell_Topological_2015, mook_spin_2016, owerreFirstTheoreticalRealization2016, nakata_magnonic_2017, nakata_Magnonic_2017a, ruckriegel_Bulk_2018},  so far there have been only a few reports on their higher-order counterparts \cite{silFirstSecondOrder2020}. This may be related to the fact that in the vicinity of sample boundaries, the magnetization field can get easily deformed, making it difficult to preserve crystalline symmetries. 

Among the two-dimensional magnetic platforms predicted to host first-order topological magnonic edge states are ferromagnetic and antiferromagnetic skyrmion crystals \cite{roldan-molina_topological_2016,diaz_topological_2019,diaz_chiral_2020}. Magnetic skyrmions are microscopic, stable, swirling spin configurations characterized by an integer topological charge \cite{nagaosa_topological_2013, fert_magnetic_2017, kanazawa_noncentrosymmetric_2017, everschor-sitte_perspective:_2018}, which in the continuum is given by $Q=\frac{1}{4\pi}\int dr^2\Bm\cdot(\partial_x\Bm\times\partial_y\Bm)$, where $\Bm = \BM/|\BM|$ is the normalized magnetization field. The strict requirements for the integer-valuedness of the net topological charge are no longer met in confined systems, thus allowing a net fractional topological charge. As a matter of fact, skyrmion nucleation has been predicted to take place from the sample edges \cite{iwasaki_current-induced_2013,du_edge-mediated_2015} through continuous growth of intermediate states with fractional topological charge. Isolated antiskyrmions, a kind of skyrmion with opposite topological charge, as well as antiskyrmion crystals were recently observed in acentric tetragonal Heusler compounds with $D_{2d}$ crystal symmetry \cite{nayak_magnetic_2017}. Furthermore, \textit{fractional antiskyrmions} stabilized along the edges of the sample have also been observed in the same compounds.

\begin{figure}[t]
\centering
\includegraphics[width=\columnwidth]{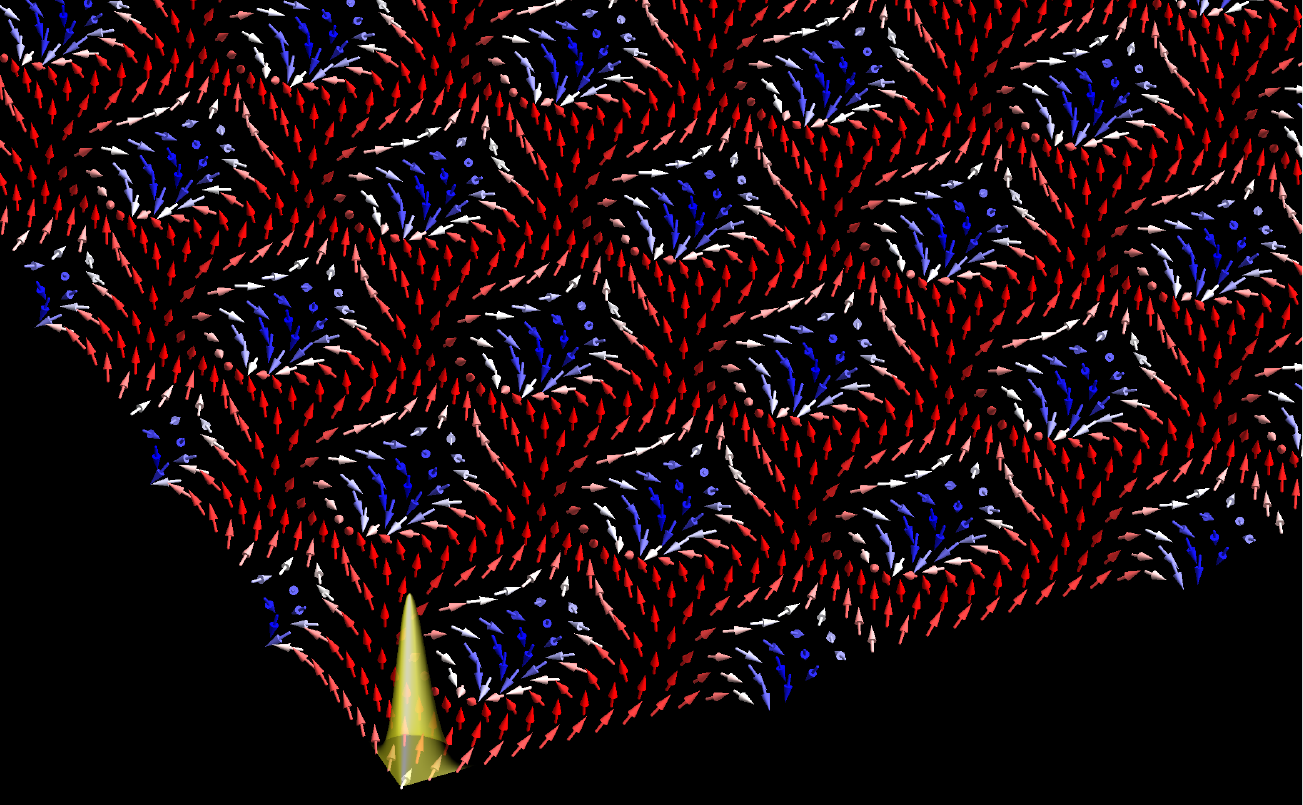}
\caption{{\bf Antiskyrmion crystals support topological magnonic corner states.} Magnetic texture of an antiskyrmion crystal in the vicinity of a sample corner. Fractional antiskyrmions that self-assemble along the sample edge allow the emergence of a topological magnonic state whose probability amplitude (depicted in green) is corner-localized.}
\label{fig1}
\end{figure}

Here, we show that a magnonic quadrupole topological insulator can be realized in two-dimensional antiskyrmion crystals, see Fig. \ref{fig1}. This second-order topological phase is protected by the combined effect of $C_{2x}\mathcal{T}$ and $C_{2y}\mathcal{T}$ symmetries, which quantize and render nontrivial the bulk quadrupole moment \cite{benalcazar_quantized_2017,benalcazar_electric_2017}. The expected robust magnonic corner states only emerge upon tuning the external magnetic field below a critical value, triggering the self-assembly of fractional antiskyrmions along the sample edges that restore the protecting symmetries, see Fig.~\ref{fig2}. Our modeling is inspired by the already available antiskyrmion-hosting Heusler compounds \cite{nayak_magnetic_2017}, in which our predictions could be experimentally tested.


\section{Antiskyrmion Crystal Model}

As a minimal model that can describe the magnetism of acentric tetragonal Heusler compounds we consider the following two-dimensional spin lattice Hamiltonian
\begin{eqnarray}\label{eq:SpinLatticeH}
H&=&\tfrac{1}{2}\sum_{\braket{\Br,\Br'}}(-J_{\Br,\Br'}\BS_{\Br}\cdot \BS_{\Br'}+ \BD_{\Br,\Br'}\cdot\BS_{\Br}\times\BS_{\Br'})\nonumber\\
&-&g\mu_\textrm{B} B_z \sum_{\Br} \BS_{\Br}\cdot\zhat \,,
\end{eqnarray}
where $\BS_{\Br}$ is a spin operator at site $\Br$ on a square lattice with lattice constant $a$. The nearest-neighbor coupling includes ferromagnetic exchange $J_{\Br,\Br'} = J( \delta_{\Br - \Br', \pm a\xhat} + \delta_{\Br - \Br', \pm a\yhat} )$ with $J > 0$, and Dzyaloshinskii-Moriya (DM) interaction $\BD_{{\Br,\Br'}} = D( \mp\xhat\delta_{\Br - \Br', \pm a\xhat} \pm\yhat\delta_{\Br - \Br', \pm a\yhat} )$ consistent with $D_{2d}$ crystal symmetry. Throughout this article we take $D/J=1.0$. The last term represents the coupling to the external magnetic field, $B_z\zhat$, where $g$ and $\mu_\textrm{B}$ denote the g-factor and Bohr magneton, respectively. 

The classical ground state texture at zero temperature is obtained using Monte Carlo simulated annealing \cite{evans_atomistic_2014} and then minimizing the energy further by solving the Landau-Lifshitz-Gilbert (LLG) equation. A triangular crystal of antiskyrmions is found in the external magnetic field range $0.27 \le g\mu_\textrm{B} B_{z}/(JS) \le 0.7$.

The natural choice for a magnetic unit cell (MUC) of the antiskyrmion crystal would span an area that contains a single antiskyrmion. Notwithstanding, it is convenient to introduce the rectangular enlarged MUC shown in Fig.~\ref{fig3}(a), which encompasses two antiskyrmions and it is commensurate with the underlying square lattice of spins. This MUC is invariant under the action of either of the symmetry operations: $C_{2z}$, twofold rotation about the $z$-axis; $C_{2x}\mathcal{T}$, twofold rotation about the $x$-axis together with time reversal; and $C_{2y}\mathcal{T}$, twofold rotation about the $y$-axis also together with time reversal. These are defined with respect to cartesian axes with origin at the center of the MUC as depicted in Fig.~\ref{fig3}(a).

Another consequence of adopting such an enlarged MUC that carries over to the spectrum of magnonic excitations is the doubling, due to backfolding, of the bulk magnon bands. For instance, Fig.~\ref{fig3}(b) shows the bottom of the bulk magnon spectrum computed using the enlarged MUC. The area highlighted in yellow corresponds to the fourth bulk magnon gap as four bands can be identified below it. Had we used a MUC comprising a single antiskyrmion, only two bands would be seen below the highlighted area, which in that case would correspond to the second bulk magnon gap. Details of the calculation of the magnon spectrum of the antiskyrmion crystal, a spatially periodic and noncollinear magnetic texture, can be found in Appendix \ref{Appendix:Edge instability}. Lastly, the double degeneracy observed for all bands along the paths $XM$ and $MX'$ is the result of and is protected by $C_{2x}\mathcal{T}$ and $C_{2y}\mathcal{T}$ symmetries (see Appendix \ref{Appendix: symmetry}).


\section{Fractional Antiskyrmions and Magnonic Corner States}

\begin{figure}[t]
\centering
\includegraphics[width=\columnwidth]{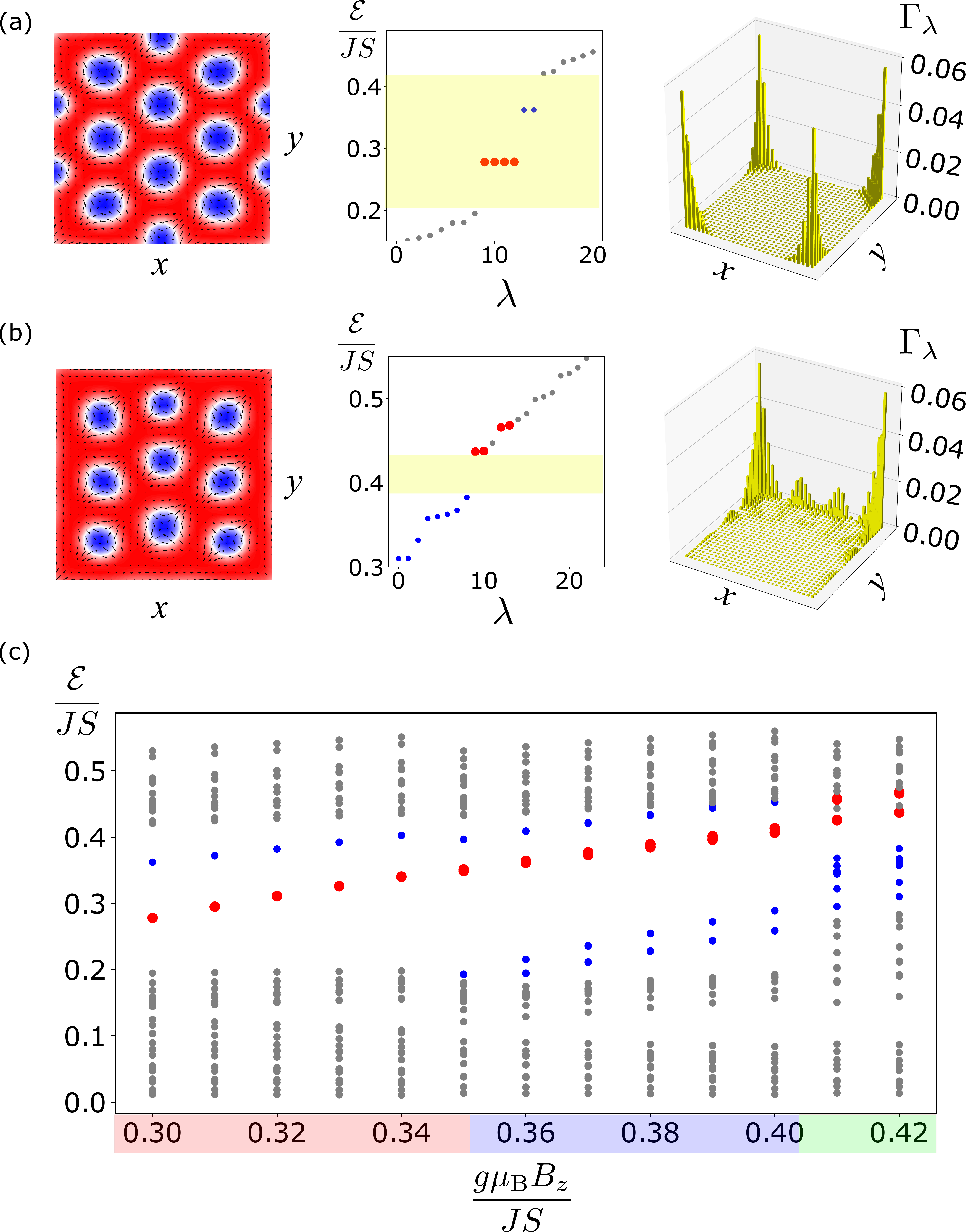}
\caption{{\bf Formation of magnonic corner states in confined antiskyrmion crystals.}
(a)-(b) Characterization of confined antiskyrmion crystals at $g\mu_\textrm{B}B_z/(JS) = 0.3$ and $g\mu_\textrm{B}B_z/(JS)= 0.42$, respectively.
Left: classical ground-state magnetic texture. Middle: magnon spectrum showing corner states (red), and trivial edge states (blue). Right: probability density of the corner states, $\Gamma_\lambda$ (defined in Appendix \ref{Appendix: hamiltonian}).
(c) Magnon spectrum against the applied magnetic field with corner and edge-localized states highlighted as in the above panels. The bottom colorbar indicates different configurations with the configuration in (a) and (b) corresponding to the red and green region, respectively. The stability and configuration of each region is studied in the Appendix \ref{Appendix:Edge instability}.
}
\label{fig2}
\end{figure}

When confined to finite-sized samples, the magnetic texture of antiskyrmions crystals exhibits a reconstruction with drastic consequences for the supported magnonic excitations. In the vicinity of the sample edges, a tendency to twist the magnetization is induced by the DM interaction which competes with the out-of-plane alignment favored by the magnetic field. For high field values, a twisted magnetic texture is attained along the edges \cite{rohart_skyrmion_2013, meynell_surface_2014}, as seen in Fig.~\ref{fig2}(b). As the field is lowered, a critical value $B_c$, with $g\mu_\textrm{B} B_{c}/(JS) = 0.41$, is reached below which the edge texture becomes unstable to the nucleation of fractional antiskyrmions (see details in Appendix \ref{Appendix:Edge instability}). Due to their mutual repulsion, antiskyrmions from within the bulk stabilize the newly nucleated fractional antiskyrmions along the edges of the sample. A similar edge instability behavior has been reported for skyrmions \cite{du_edge-mediated_2015,muller_edge_2016}, but so far no edge-stabilized fractional skyrmions have been observed. 

The magnon spectrum of a confined antiskyrmion crystal (see details in Appendix \ref{Appendix: hamiltonian}) at $g\mu_\textrm{B} B_{z}/(JS) = 0.32$ with fractional antiskyrmions stabilized along the sample edge is shown in Fig.~\ref{fig2}(a). Highlighted in red are four degenerate states which are well separated from the bulk states (gray). They correspond to corner states, one for each corner of the sample. States highlighted in blue right above the corner states, are local excitations of fractional antiskyrmions. On the other hand, at $g\mu_\textrm{B} B_{z}/(JS) = 0.42$, there are no fractional antiskyrmions [see Fig.~\ref{fig2}(b)]. Although four modes with significant probability density near the corners can still be identified, they are buried among bulk modes and they spread over the edges and into the bulk of the sample.

A clearer picture emerges from plotting the magnon spectrum as a function of the applied magnetic field, shown in Fig.~\ref{fig2}(c). We first note that the bulk magnon gap where corner states are found corresponds to the fourth bulk magnon gap of the antiskyrmion crystal, shown in Fig.~\ref{fig3}(b). Remarkably, corner states emerge only when fractional skyrmions are stabilized along the sample edges, i.e., for $g\mu_\textrm{B} B_{z}/(JS) \leq 0.41$. In this range, the energies of corner states (red) and states localized at fractional antiskyrmions (blue) increase linearly with applied field. 

These magnonic corner states are remarkably robust against the effect of magnetic impurity disorder. Even though their degeneracy is lifted by disorder, they remain isolated avoiding hybridization with bulk modes. Details of our disorder study can be found in Appendix \ref{Appendix: disorder}.


\section{Bulk Quadrupole Moment}

\begin{figure}[t]
\centering
\includegraphics[width=\columnwidth]{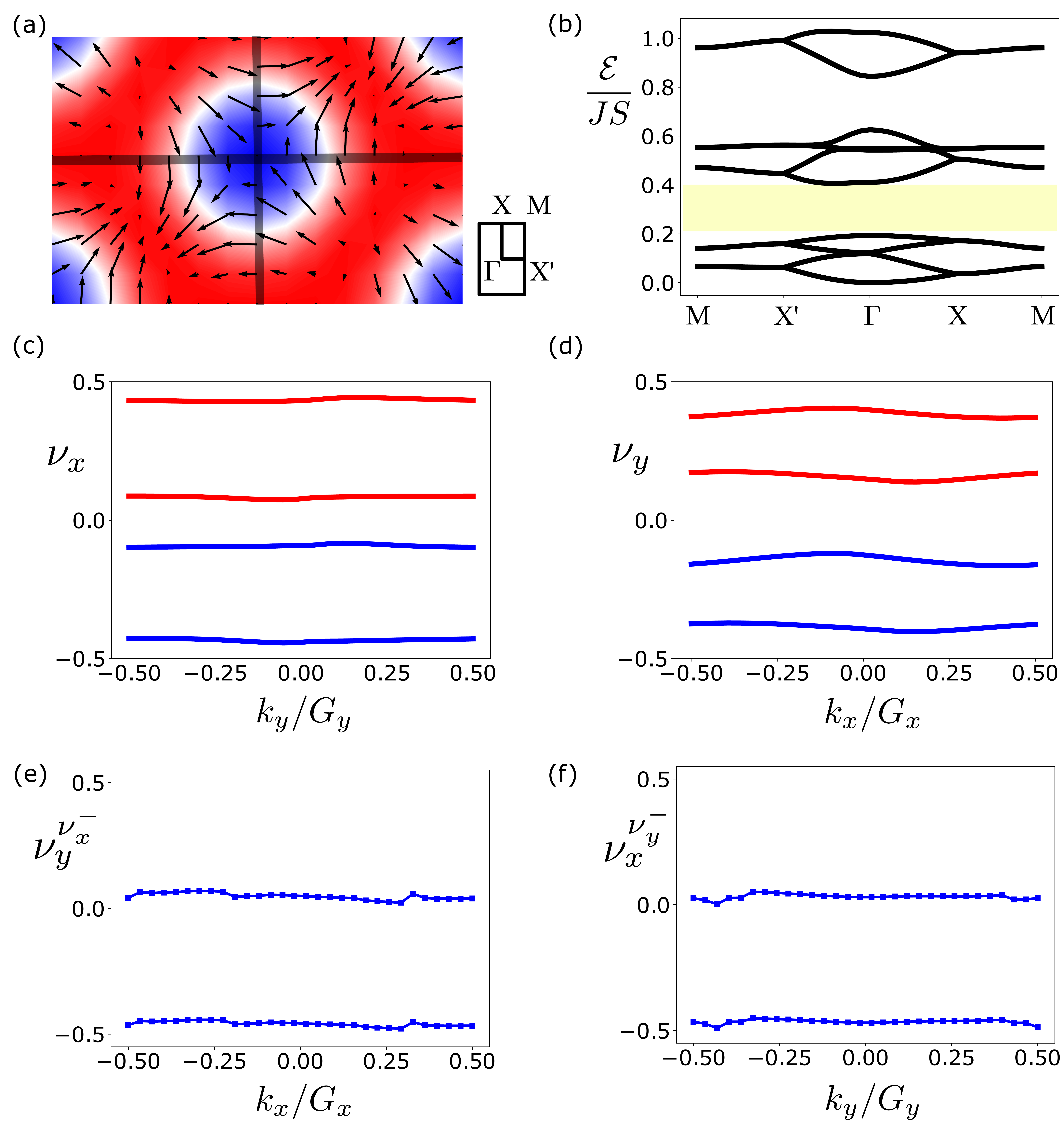}
\caption{{\bf Bulk symmetries and Wannier spectra in antiskyrmion crystals.} 
(a) Magnetic unit cell of the antiskyrmion crystal with symmetry lines for $C_{2x/2y}$. The $C_{2z}$ rotation axis is at the center of the magnetic unit cell. The first Brillouin zone is also shown. 
(b) Bulk magnon spectrum of the antiskyrmion crystal. The fourth bulk magnon gap is highlighted in yellow.
(c)-(d) Wannier spectrum of the four lowest-energy magnon bands showing the Wannier sector  $\nu_{x/y}^{+}$ in red and $\nu_{x/y}^-$ in blue.
(e)-(f) Wannier centers of the Wannier sector $\nu_{x/y}^{-}$. For all panels, the magnetic field is $g\mu_\textrm{B}B_z/(JS)= 0.35$ and $G_{x/y}=2\pi/(L_{x/y}a)$ with $L_{x/y}$ denoting the number of lattice sites in the magnetic unit cell along $x/y$-axis. 
}
\label{fig3}
\end{figure}

\begin{figure*}[t]
\centering
\includegraphics[width=1.6\columnwidth]{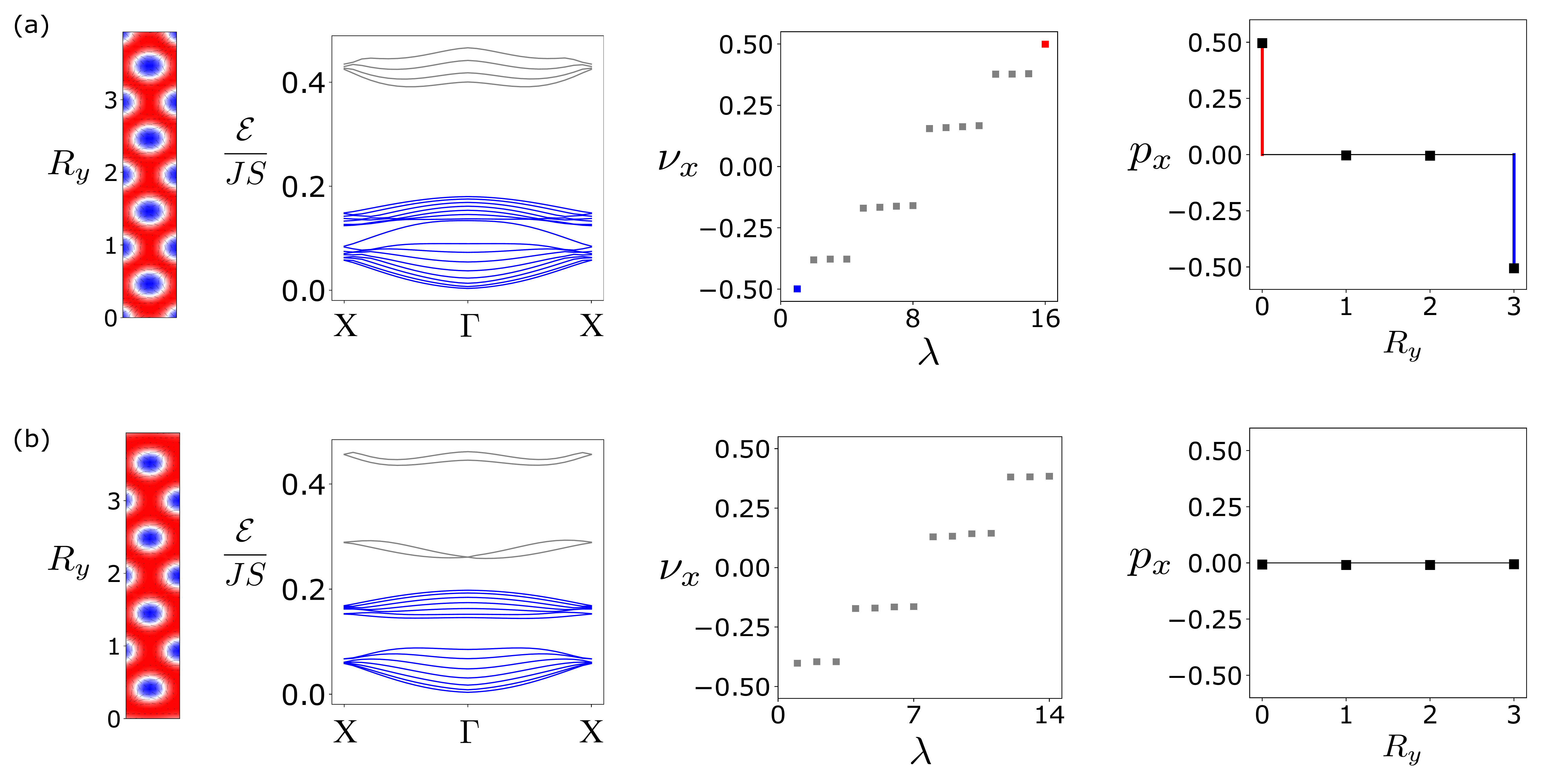}
\caption{{\bf Edge dipole moment in a strip geometry.} 
Studies of the edge dipole moment (a) below ($g\mu_\textrm{B}B_z/(JS)= 0.3$) and (b) above ($g\mu_\textrm{B}B_z/(JS)= 0.41$) the edge instability.
Column 1: Magnetic unit cell of the antiskyrmion crystal in a strip geometry, which extends infinitely along the $x$-axis (horizontal) and whose width fits four magnetic unit cells of the bulk antiskyrmion crystal (vertical). The vertical axis on the left indicates the position of the bulk antiskyrmion crystal magnetic unit cell along the $y$-axis.
Column 2: Magnon spectrum of the antiskyrmion crystal in a strip geometry with bulk bands below the second gap indicated in blue.
Column 3: Wannier spectrum $\nu_x$ showing four bulk bands (grey) and the edge modes (red and blue). 
Column 4: Polarization $p_x(R_y)$ averaged over each vertical unit cell.}
\label{fig4}
\end{figure*}

The robust magnonic corner states we have uncovered are in fact second-order topological magnonic states. More precisely, antiskyrmion crystals provide a platform to realize magnonic quadrupole topological insulators. This higher-order topological phase is protected by the combined action of $C_{2x}\mathcal{T}$ and $C_{2y}\mathcal{T}$ symmetries, resulting in a quantized bulk quadrupole moment.

We use the nested Wilson loop construction (for details, see Appendix \ref{Appendix: nested}) to compute the bulk quadrupole moment. However, before doing so, the following three essential requirements must be fulfilled \cite{benalcazar_electric_2017}: (1) the corner states lie within the $n$-th bulk magnon gap with $n \geq 2$; (2) the lowest $n$ magnon bands carry a vanishing net Chern number; and (3) the bulk dipole moment vanishes.

As already mentioned above, the magnonic corner states are found within the fourth bulk magnon gap. Also, the four bulk magnon bands below this gap carry no Chern number in the magnetic field range of interest (which is also true for the ferromagnetic skyrmion crystal). The cartesian components of the bulk polarization are given by 
\begin{align}
p_x(k_y) &= \sum_{j=1}^M  \nu_x^j(k_y)  \textrm{ mod 1} \,, \\
p_y(k_x) &= \sum_{j=1}^M  \nu_y^j(k_x)  \textrm{ mod 1} \,,
\end{align}
where $\nu_x^j(k_y)$ and $\nu_y^j(k_x)$ are Wannier centers of bulk magnon bands (see appendix \ref{Appendix: wilson loop}) and $M = 4$ is the number of bulk bands below the gap where the magnonic corner states are found. It can be shown that the symmetries of the MUC translate into the following constraints on the Wannier spectra
\begin{align}
\nu_x^j(k_y) &\overset{C_{2y}\mathcal{T}}{=} -\nu_x^j(-k_y) \textrm{ mod }1 \,, \\
\nu_y^j(k_x) &\overset{C_{2x}\mathcal{T}}{=} -\nu_y^j(-k_x) \textrm{ mod }1 \,.
\end{align}
Therefore, $\nu_x^j$ can be constant and equal to either $0$ or $\frac{1}{2}$, the other possibility being that it has a partner $\nu_x^{j'}$ such that $\nu_x^j(k_y) = - \nu_x^{j'}(-k_y)$. Similar allowed values can be expected for $\nu_y^j$. In Fig.~\ref{fig3}(c)-(d) show the Wannier spectra of the lowest four bulk magnon bands. The absence of flat Wannier bands at $\frac{1}{2}$ implies that the bulk dipole moment vanishes, i.e., $(p_x, p_y) = (0, 0)$.

Now that we have established the fulfillment of the above three desiderata, we proceed to compute the bulk quadrupole moment. We refer the interested reader to Appendix \ref{Appendix: nested} where details of this calculation can be found.  The bulk quadrupole moment is given by
\begin{align}
q_{xy}=2p_y^{\nu_x^-}p_x^{\nu_y^-} \,,
\end{align}
where $p_x^{\nu_y^-}$ and $p_y^{\nu_x^-}$ are Wannier sector polarizations. In turn, these are respectively given by sums over the Wannier sector bands $\nu_{y}^{\nu_x^-,p}(k_x)$ and $\nu_{x}^{\nu_y^-,p}(k_y)$, defined in Eq.~\eqref{nested_center}. The two Wannier sector bands expected for the present construction are shown in Fig.~\ref{fig3}(e)-(f). Within numerical error, one of them is quantized to $0$ and the other to $-\frac{1}{2}$, as constrained by $C_{2x}\mathcal{T}$ and $C_{2y}\mathcal{T}$ symmetries. Therefore, we obtain $p_x^{\nu_y^-} = p_y^{\nu_x^-} = -\frac{1}{2}$, which imply the quantization of the bulk quadrupole moment to $q_{xy} = \frac{1}{2}$.

Furthermore, our numerical calculations show that the bulk quadrupole moment is quantized for any value of the external magnetic field, as long as the antiskyrmion crystal remains stable.


\section{Bulk-Boundary Correspondence and Fractional Antiskyrmions}

The emergence of magnonic corner states is not guaranteed by a quantized bulk quadrupole moment. Boundaries must also preserve the protecting $C_{2x}\mathcal{T}$ and $C_{2y}\mathcal{T}$ symmetries. At high applied fields, even though the bulk quadrupole moment is quantized, no magnonic corner states are realized because the magnetic texture of the antiskyrmion crystal is distorted near the sample boundaries [see Fig.~\ref{fig2}(b)], thus breaking the protecting symmetries. Decreasing the magnetic field below the critical value $B_c$ triggers the nucleation of fractional antiskyrmions from the sample boundaries. The newly formed fractional antiskyrmions play the role of the nearest neighbors missing from antiskyrmions located near the edge of the sample. Therefore, by virtue of their mutual repulsion with bulk antiskyrmions, fractional antiskyrmions self-assemble along the sample edges. This process restores the protecting symmetries, thus allowing the formation of magnonic corner states.

Two hallmark signatures are expected of a quantized quadrupole moment: edge dipole moments and corner charges. These should also be quantized in a manner consistent with the bulk quadrupole, i.e., $q_{xy} = |p_x^{\rm edge}| = |p_y^{\rm edge}| = |Q_c|$. The edges of the sample are themselves topological insulating and the corner states are simultaneous end states of two converging edges \cite{benalcazar_electric_2017,benalcazar_quantized_2017}. Therefore, as an initial consistency check we can just count the number of magnonic corner states. Only one such state is expected at each corner.

In order to compute the edge dipole moments we study the antiskyrmion crystal on a strip geometry (see Appendix \ref{Appendix:Edge} for details of this calculation). The results are reported in Fig.~\ref{fig4} for two values of the applied magnetic field, above and below $B_c$. The MUC of the strip (leftmost panels) is periodic along the horizontal, $x$-axis and fits four MUC of the bulk antiskyrmion crystal along its width. When the applied field is below $B_c$ [Fig.~\ref{fig4}(a)], fractional antiskyrmions stabilize along the top and bottom edges. The blue bands from the one-dimensional magnon spectrum, which lie below the gap where magnonic corner states emerge, are used in the calculation of the Wannier spectrum $\nu_x$. The polarization $p_x(R_y)$ indicates the presence of quantized edge dipole moments with opposite values at opposite edges: $\frac{1}{2}$/$-\frac{1}{2}$ at the top/bottom edge. Crucially, the edge Wannier states are localized at the core of fractional skyrmions along the edges. On the other hand, for an applied field larger than $B_c$ [Fig.~\ref{fig4}(b)], the strip edges do not host fractional antiskyrmions and the polarization vanishes throughout the width of the strip. It is important to note that in this case, trivial magnonic edge states appear within the magnon gap of interest. These should not be included in the Wannier spectrum calculation. We also confirmed that $p_y(R_x)$ shows a similar behavior above and below the critical field.

A precise definition of the corner charge \cite{park_fractional_2016, thakurathi_fractional_2018, pletyukhov_topological_2020, pletyukhov_surface_2020} for magnonic systems as well as details of its calculation are given in Appendix \ref{Appendix: Qc}. We find that each of the four corners hosts a quantized corner charge that satisfies $|Q_c| = \frac{1}{2}$ (see Fig.~\ref{fig12}). It should be pointed out that the specific configuration of the corner charges depends on how the degeneracy of the magnonic corner states is broken \cite{benalcazar_electric_2017}.


\section{Engineering Corner States}

\begin{figure}[t]
\includegraphics[width=70mm]{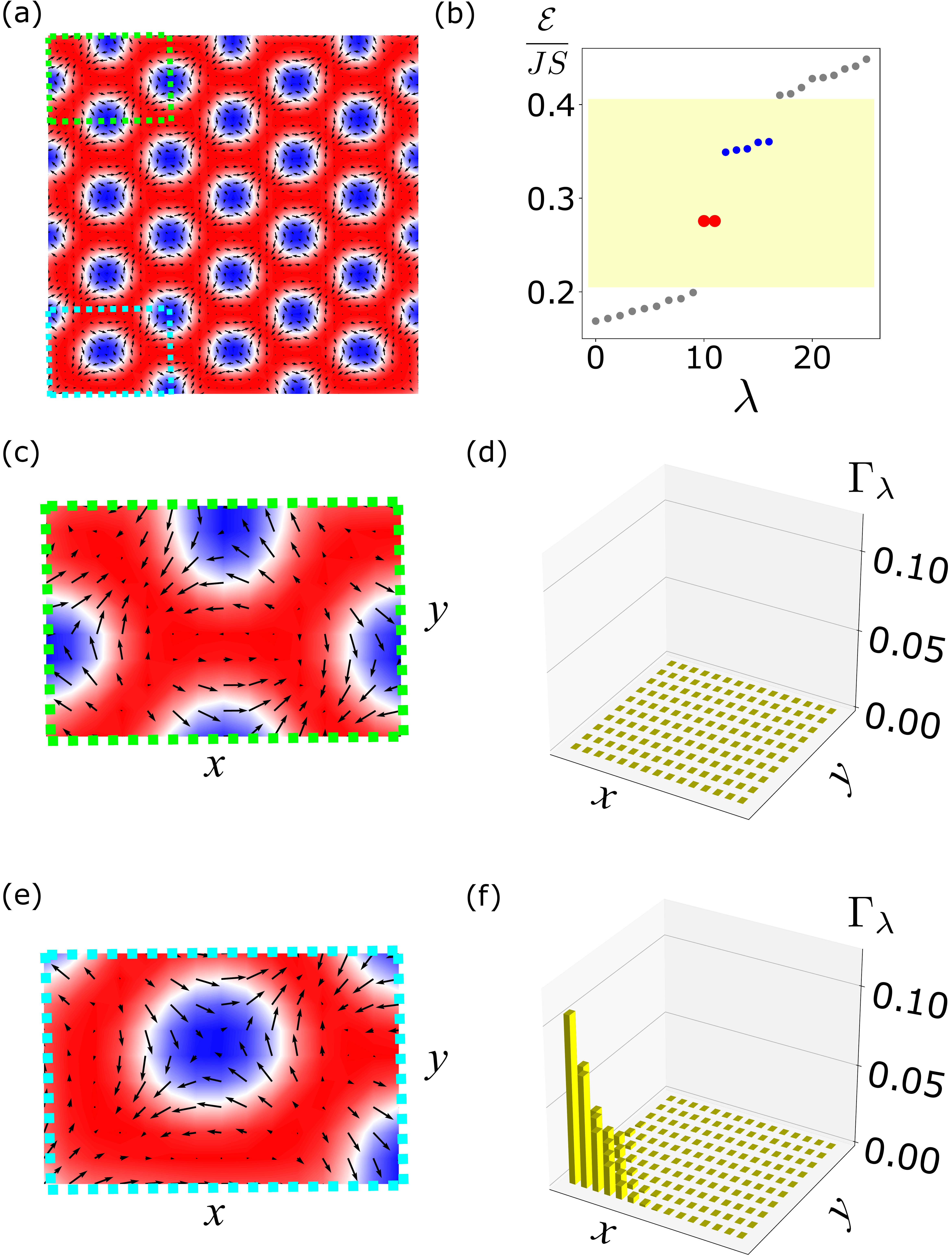}
\centering
\caption{ \textbf{Engineering magnonic corner states at select corners.}
(a) Classical ground-state texture of a confined antiskyrmion crystal. The top (bottom) left corner is enclosed by a green (light blue) dashed rectangle. (b) Magnon spectrum showing two degenerate corner states (red) and several edge-localized states (blue). The top and bottom corners enclosed in (a) are enlarged in (c) and (e), respectively. Fractional antiskyrmions are further away from the bottom than the top corners. (d), (f) Probability density of the corner states plotted in the vicinity of the corresponding corner (c), (e). Only the bottom corners host corner states.
}
\label{fig5}
\end{figure}

Below the critical field of the edge instability, fourfold degenerate corner states are obtained robustly for a sufficiently large system as the antiskyrmion crystal can deform slightly to adapt to a possible mismatch between the magnetic unit cell and the sample size. However, finite size effects become important when the magnetic unit cell and sample have comparable sizes. This leads to an additional criterion for the formation of magnonic corner states.

In Fig.~\ref{fig5}, we show the classical ground-state texture of a $45\times 45$ spin lattice system, fitting 4.5 magnetic unit cells along the vertical direction. As a result, the local arrangement of fractional antiskyrmions is no longer equivalent between top and bottom corners. Figures ~\ref{fig5}(c) and \ref{fig5}(e) show that fractional antiskyrmions are stabilized closer to the top than to the bottom corners. Although the configuration of Fig.~\ref{fig5}(e) is energetically more favorable due to the repulsion between fractional antiskyrmions in Fig.~\ref{fig5}(c), the mismatch between the magnetic unit cell and system size enforces the energetically less favorable configuration near the top corners. The local arrangement of fractional antiskyrmions near the sample corners plays an important role in the formation of magnonic corner states. As revealed by the probability density in Figs.~\ref{fig5}(d) and \ref{fig5}(f), magnonic corner states are supported only at the bottom corners. This observation is explained by the seam dependence of the Wannier centers and polarization as discussed in Appendix \ref{Appendix:Edge} \cite{watanabe_inequivalent_2018}.
Therefore, adjusting the size of the sample is a possible route towards controlling the location of magnonic corner states.

\begin{figure}[t]
\centering
\includegraphics[width=\columnwidth]{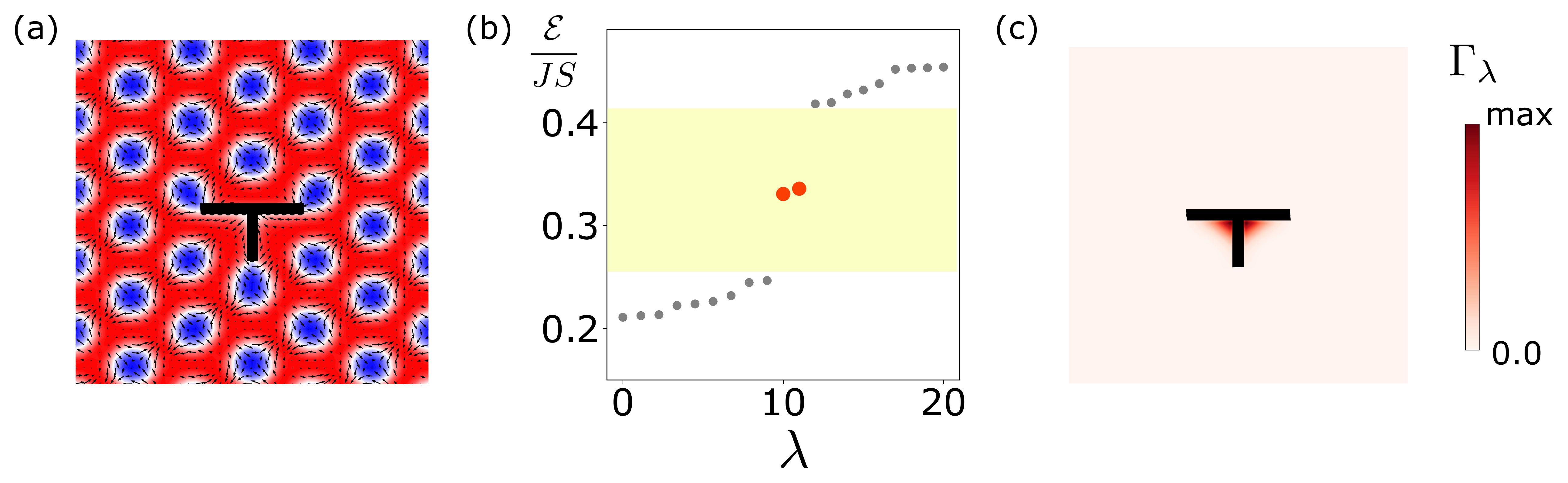}
\caption{{\bf Engineering magnonic corner states inside the sample.} (a) Classical ground-state magnetic texture of a $45\times 45$ periodic system with a $T$-shaped hole at the center of the sample at  $g\mu_\textrm{B}B_z/(JS)= 0.31$. (b) Magnon spectrum at the $\Gamma$ point showing corner states (red) within the fourth bulk magnon gap (yellow). (c) Probability density of the corner states localized at the inner corners of the $T$-shaped hole.}
\label{fig6}
\end{figure}

It would be highly desirable to have an alternative means to engineer the location of magnonic corner states without modifying the sample size. This has led us to consider holes within the bulk of the sample. Figure~\ref{fig6}(a) shows the spin configuration induced by a T-shaped hole at $g\mu_\textrm{B} B_{z}/(JS)=0.31$. The system consists of $45 \times 45$ sites with periodic boundary conditions to model an infinitely large sample. By diagonalizing the spin wave Hamiltonian, we obtain in-gap states that localize at the inner corners of the T-shaped hole [see Figs.~\ref{fig6}(b) and \ref{fig6}(c)], as long as the symmetries that protect the bulk quadrupole moment are not locally broken. This convenient approach could be extended to engineer magnonic corner states across the sample.

\section{Discussion}

In addition to a quantized bulk quadrupole moment, the formation of fractional antiskyrmions is essential to obtain topologically protected magnonic corner states. While fractional antiskyrmions have been observed in the acentric tetragonal Heusler compound Mn$_{1.4}$Pt$_{0.9}$Pd$_{0.1}$Sn~\cite{nayak_magnetic_2017}, no counterpart has been observed for ferromagnetic Bloch or N\'{e}el skyrmions \cite{du_edge-mediated_2015, song_quantification_2018}. Our numerical simulations indicate that the presence of the (anti)skyrmion crystal is of paramount importance for the stabilization of fractional (anti)skyrmions along the edges of the sample. Therefore, we suggest using a field-cooling protocol for the experimental observation of fractional skyrmions.

Here we have focused on the experimentally available antiskyrmion crystals stabilized by $D_{2d}$ crystalline symmetry. However, our findings generalize to ferromagnetic Bloch skyrmion crystals, since they are also
$C_{2x}\mathcal{T}$- and $C_{2y}\mathcal{T}$-symmetric. Within the same theoretical framework, but using $M_x\mathcal{T}$ and $M_y\mathcal{T}$ instead of $C_{2x}\mathcal{T}$ and $C_{2y}\mathcal{T}$ as the protecting symmetries, our results could also apply to ferromagnetic N\'{e}el skyrmion crystals.

In this work, we have studied a quadratic spin wave Hamiltonian on the basis of the linear spin wave approximation, which neglects magnon-magnon interactions. A couple of comments on the stability of the magnonic quadrupole topological insulating phase are in order. Firstly, the robustness of magnonic corner states was confirmed by micromagnetic simulations using the LLG equation (see Appendix \ref{Appendix:LLG simulation}), which contains all orders of nonlinear terms in the classical limit~\cite{mook_spin_2016}. Furthermore, it has been recently shown that the many-body effects of magnon-magnon interactions do not cause spontaneous quasiparticle decay of the lowest four magnon bands of skyrmion crystals \cite{mook_quantum_2020}, which are needed in the construction of the bulk quadrupole moment. Based on the symmetry considerations discussed above, we expect this result to also extend to antiskyrmion crystals. Therefore, the higher-order topological magnonic corner states supported by antiskyrmion crystals should be robust even beyond linear spin wave theory.


\section{Conclusions}

We uncover that antiskyrmion crystals can realize a magnonic quadrupole topological insulator. Tuning an applied magnetic field induces the self-assembly of fractional antiskyrmions along the edges of the sample. Remarkably, these fractional antiskyrmions restore the protecting symmetries that allow the formation of robust magnonic corner states. Acentric tetragonal Heusler compounds, where antiskyrmion crystals have already been observed, constitute an ideal platform to test our findings. Magnonic corner states can be used as a magnon cavity \cite{hans_high_2013} with a high Q factor \cite{chen_direct_2019} to enhance magnon-photon \cite{Tabuchi2015,Li2019} interactions for quantum computing and quantum information applications. Our study highlights a new form of topological excitations in magnetic systems and its potential use in the design of future magnonic devices.


\begin{acknowledgments}
We are grateful to A. Mook, I. Tateishi, T. Hinokihara, H. Matsuura, M. Ogata, M. Nagao, T. Nagase, D. Shimizu for useful discussions. T.H. is supported by Japan Society for the Promotion of Science through Program for Leading Graduate Schools (MERIT) and JSPS KAKENHI (Grant No. 18J21985). This work was supported by the Swiss National Science Foundation and NCCR QSIT. This project received funding from the European Union’s Horizon 2020 research and innovation program (ERC Starting Grant, Grant Agreement No. 757725).
\end{acknowledgments}


\appendix


\section{Spin wave Hamiltonian}
\label{Appendix: hamiltonian}

Quantum fluctuations about classical ground-state spin textures are treated by solving the spin wave Hamiltonian. For this purpose, we employ the Holstein-Primakoff (HP) transformation \cite{holstein_field_1940}. 
It is convenient to define the spin operators with respect to the local orthonormal basis $(\Be_{\Br}^{1},\Be_{\Br}^{2},\Bm_{\Br})$, where $\Bm_{\Br}$ is parallel to the ground-state spin texture and $\Be_{\Br}^{1}\times\Be_{\Br}^{2} = \Bm_{\Br}$. In this local basis, the spin operators read $\bs{S}_{\Br} = \SS_{\Br}^{1}\Be_{\Br}^{1} + \SS_{\Br}^{2}\Be_{\Br}^{2} + \SS_{\Br}^{3}\Bm_{\Br}$. The HP transformation is performed by substituting $\SS_{\Br}^{+} = (2S - a^\dagger_{\Br} a_{\Br})^{\frac{1}{2}} a_{\Br}$, $\SS_{\Br}^{-} = a_{\Br}^\dagger(2S - a^\dagger_{\Br} a_{\Br})^{\frac{1}{2}}$, and $\SS_{\Br}^{3} = S - a_{\Br}^\dagger a_{\Br}$, where $\SS_{\Br}^{\pm} = \SS_{\Br}^{1} \pm i\SS_{\Br}^{2}$, and $a_{\Br}$, $a_{\Br}^\dag$ are the HP bosonic operators. Assuming $S\gg 1$, the Hamiltonian is expanded as a series in $1/S$. The free spin wave Hamiltonian is constructed by collecting terms quadratic in the HP operators
\begin{align}
\Hsw = \frac{S}{2}\sum_{\Br,\Br'}\psi_{\Br}^\dagger H_{\Br,\Br'} \psi_{\Br'}+\CalE_0 \,,
\end{align}
where $\psi_{\Br}=(a_{\Br},a^\dagger_{\Br})^T$, $\CalE_0=-\frac{1}{2}S\sum_{\Br} \Lambda_{\Br}$ and
\begin{align}
H_{\Br,\Br'} =
\begin{pmatrix}
\Omega_{\Br,\Br'} & \Delta_{\Br,\Br'}\\
\Delta_{\Br,\Br'}^* &\Omega_{\Br,\Br'}^* 
\end{pmatrix}\,.
\end{align}
Each expression is given by $\Omega_{\Br,\Br'} = \delta_{\Br,\Br'}\Lambda_{\Br} + \frac{1}{2}[ -J_{\Br,\Br'}\Be_{\Br}^{+}\cdot\Be_{\Br'}^{-} + \BD_{\Br,\Br'}\cdot\Be_{\Br}^{+}\times\Be_{\Br'}^{-} ]$, $\Delta_{\Br,\Br'} = \frac{1}{2}[ -J_{\Br,\Br'}\Be_{\Br}^{+}\cdot\Be_{\Br'}^{+} + \BD_{\Br,\Br'}\cdot\Be_{\Br}^{+}\times\Be_{\Br'}^{+} ]$, and $\Lambda_{\Br} = \sum_{\Br'}[ J_{\Br,\Br'}\Bm_{\Br}\cdot\Bm_{\Br'} - \BD_{\Br,\Br'}\cdot\Bm_{\Br}\times\Bm_{\Br'} ] + \frac{g\mu_\textrm{B}B_{z}}{S}\zhat\cdot\Bm_{\Br}$, with $\Be_{\Br}^{\pm} = \Be_{\Br}^{1} \pm i\Be_{\Br}^{2}$.

The magnon spectrum is obtained by diagonalizing the spin wave Hamiltonian with a paraunitary matrix $T_{\Br}$, which satisfies $T_{\Br}^\dagger \Sigma T_{\Br}=T_{\Br} \Sigma T_{\Br}^\dagger=\Sigma$ where 
\begin{align}\label{eq:Sigma}
\Sigma = 
\begin{pmatrix}
\mathds{1}_{N\times N} & 0 \\
0 & - \mathds{1}_{N\times N} \\
\end{pmatrix} \,,
\end{align}
with $\mathds{1}_{N\times N}$ being the identity matrix of order $N$, and $N$ is the total number of sites. The diagonalized spin wave Hamiltonian is given by
\begin{align}
\Hsw = S\sum_{\lambda} \CalE_\lambda \big(\alpha_\lambda^\dagger\alpha_\lambda+\tfrac{1}{2} \big)+\CalE_0 \,,
\end{align}
where $\lambda$ is the index for each magnon mode, $\CalE_\lambda$ is the corresponding eigenvalue, and $(\alpha_\lambda,\alpha_\lambda^\dagger)^T=T_{\Br}(a_{\Br},a_{\Br}^\dagger)^T$.

For a given a spin texture, we characterize the spatial distribution of its $\lambda$-th magnon mode by the corresponding magnonic probability density
\begin{align}
\Gamma_\lambda(\Br) = |\bra{GS}a_{\Br}\alpha_\lambda^\dagger \ket{GS}|^2 \,,
\end{align}
where $\ket{GS}$ is the vacuum state of magnons, i.e., $\alpha_\lambda \ket{GS} = 0$.

When the ground-state spin texture is spatially periodic, as is the case of antiskyrmion crystals in bulk, crystal momentum $\Bk$ can be introduced. As described in detail in Refs. \cite{diaz_topological_2019,diaz_chiral_2020}, the spin wave Hamiltonian can be written in reciprocal space as
\begin{align}
\Hsw = \frac{S}{2}\sum_{\Bk}\psi_{\Bk i}^\dagger H_{\Bk}^{ij} \psi_{\Bk j}+\mathcal{E}_0 \,,
\end{align}
where $\psi_{\Bk i}=(a_{\Bk i}, a^\dagger_{-\Bk i})^T$ with $i$ labeling the $\tilde{N}$ spins that comprise the magnetic unit cell. In this case, the spin wave Hamiltonian is diagonalized by a paraunitary matrix $T_{\Bk}$ satisfying $T_{\Bk}^\dagger \sigma_3 T_{\Bk} = T_{\Bk} \sigma_3 T_{\Bk}^\dagger = \sigma_3$, where
\begin{align}\label{eq:sigma3}
\sigma_3 = 
\begin{pmatrix}
\mathds{1}_{\tilde{N}\times\tilde{N}} & 0 \\
0 & - \mathds{1}_{\tilde{N}\times\tilde{N}} \\
\end{pmatrix} \,,
\end{align}
We use the  numerical diagonalization method described in Ref.~\cite{colpa_diagonalization_1978} to compute the magnon spectra, paraunitary matrices, and magnonic probability densities.


\section{Micromagnetic simulations}
\label{Appendix:LLG equation}

Substituting the spin operators $\BS_{\Br}$ in the spin lattice Hamiltonian \eqref{eq:SpinLatticeH} by $S\Bm_{\Br}$, where $\Bm_{\Br}$ is a unit vector, we get a classical magnetic energy function. After this substitution, we can describe classical magnetization dynamics of the lattice systems we consider here using the LLG equation,
\begin{align}
\frac{d\Bm_{\Br}}{dt}=-\frac{\gamma \Bm_{\Br}}{1+\alpha^2}\times\left[ \BH_{\Br}^{\rm{eff}} + \alpha\Bm_{\Br}\times\BH_{\Br}^{\rm{eff}} \right] \,, 
\end{align}
where $\BH_{\Br}^{\rm{eff}} = -[1/(\hbar \gamma S)]\partial H/\partial \Bm_{\Br}$, $\gamma = g\mu_\textrm{B}/\hbar$ is the gyromagnetic ratio, and $\alpha$ is the Gilbert damping constant. Throughout this article we use $\alpha=0.04$. Time is measured in units of $\hbar/(JS)$. Assuming the exchange coupling $J=1$ meV, and $S = 1$ the time scale is in the order of $0.7$ ps.


\section{Edge instability and fractional antiskyrmions}
\label{Appendix:Edge instability}

\begin{figure}[t]
\centering
\includegraphics[width=0.8\columnwidth]{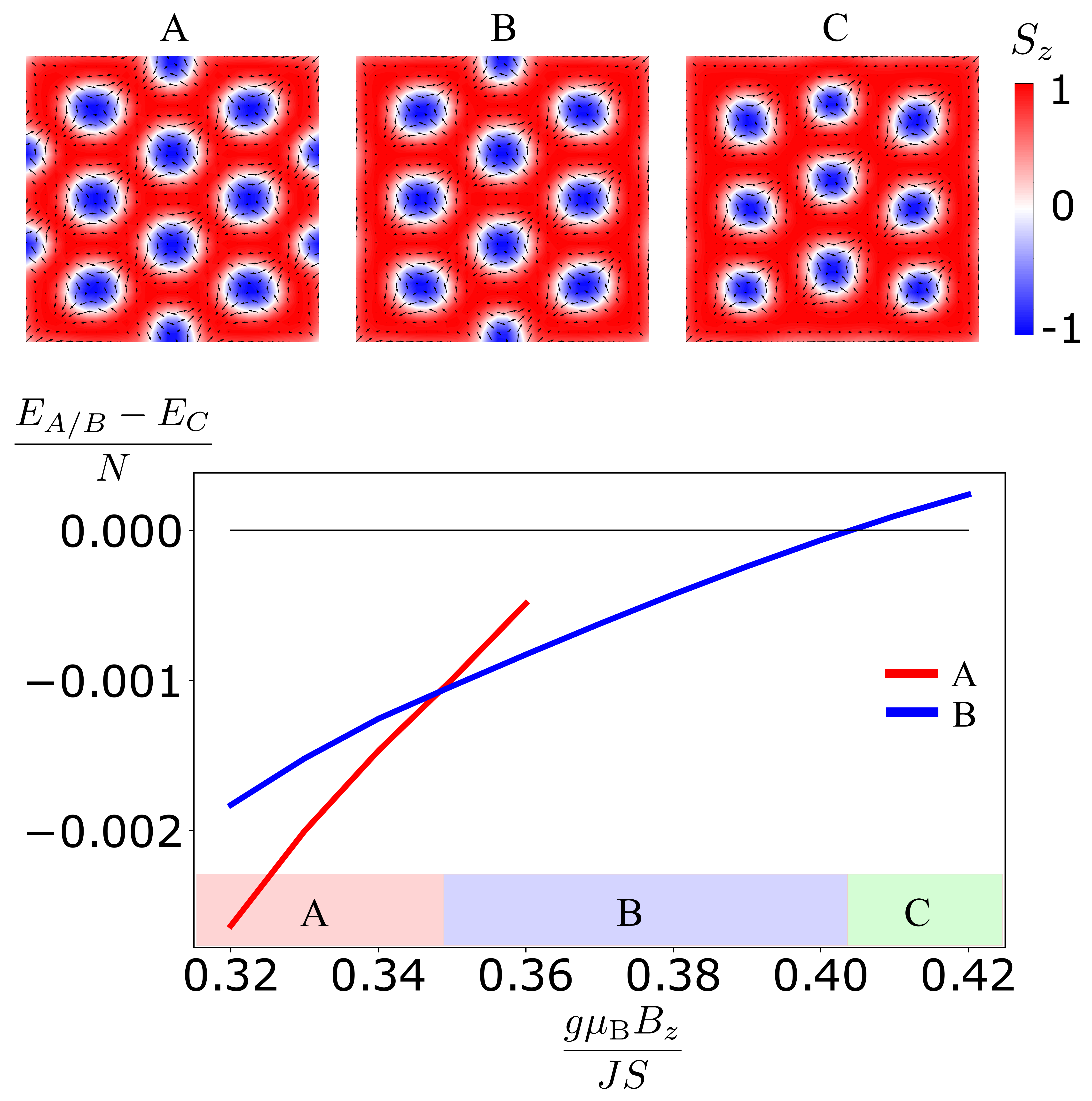}
\caption{\textbf{Fractional antiskyrmions stabilized along the sample edge.}
Top: Classical ground-state magnetic textures of confined antiskyrmion crystals for different applied magnetic field values, representative of the possible edge configurations. Although absent at high fields (C), fractional antiskyrmions are progressively stabilized along the edge as the field is lowered (B, A). 
Bottom: Energies of configurations A, B, and C  as functions of the applied magnetic field. As the field decreases, the minimum energy configuration changes from C to B to A. 
}
\label{fig7}
\end{figure}

The magnetic texture in materials that admit either bulk or interfacial DM interactions acquires a twist near the sample edges \cite{Wilson2013,Rohart2013,meynell_surface_2014}. Furthermore, the twisted magnetic texture becomes unstable at a critical magnetic field obtained from a continuum field theory of topologically trivial bound magnon edge modes \cite{muller_edge_2016}. Below this critical field, stripe domains can grow from the edges into a field-polarized bulk.

A similar behavior is expected to occur in magnetic materials with $D_{2d}$ DM interaction. In order to estimate the corresponding critical field, we simulate the classical time evolution of a finite system of $30\times 30$ spins with open boundary conditions employing the LLG equation. We first prepare a field-polarized phase at $g\mu_\textrm{B}B_z/(JS)=0.8$ via Monte Carlo simulated annealing, which is taken as the initial state of the LLG equation. The magnetic field is then gradually reduced until we observe the creation of stripe domains from sample edges. As a stripe domain grows from the edge, it elongates into the sample. We find that the edges become unstable to the creation of stripe domains below $g\mu_\textrm{B}B_z/(JS)=0.41$, in agreement with Ref. \cite{muller_edge_2016}. In contrast, when the initial state is a confined antiskyrmion crystal, the elongation of nascent stripe domains from edges into the bulk is hindered by the mutual repulsion with antiskyrmions within the sample. This leads to antiskyrmions with fractional charge located along the edges of the sample .

In order to study the stability of these fractional antiskyrmions, we compute the classical ground-state texture
for a spin lattice of $30\times 30$ sites with open boundary conditions for a range of applied magnetic field using the LLG equation. In Fig.~\ref{fig7} we identify three different configurations and plot their average energy per site at different magnetic fields. We find that fractional antiskyrmions are stabilized below the critical field and become stable on all edges at $g\mu_\textrm{B}B_z/(JS)\approx0.35$. Note that the topologically protected magnonic corner states are obtained in both configurations A and B as shown in Fig.~\ref{fig2}(c).


\section{Excitation of magnonic corner states by an AC magnetic field}
\label{Appendix:LLG simulation}

\begin{figure}[t]
\centering
\includegraphics[width=0.8\columnwidth]{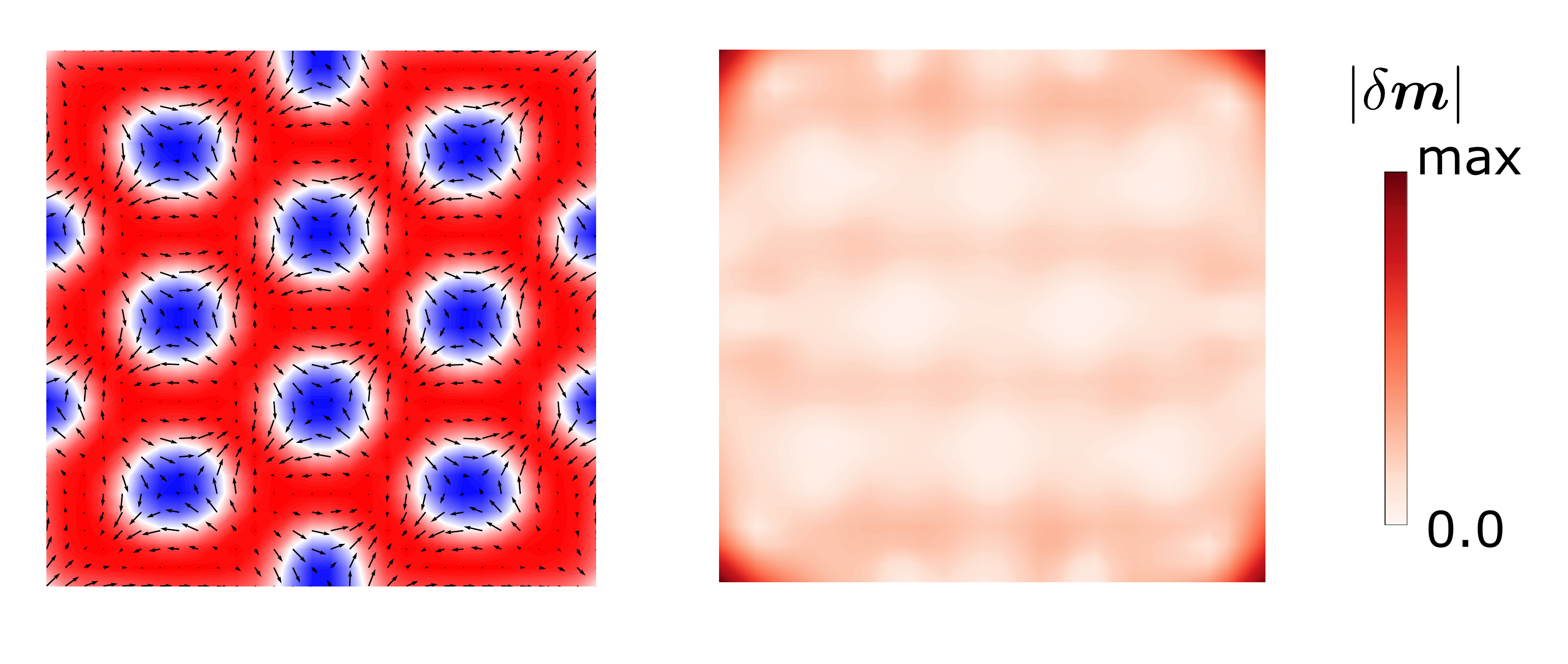}
\caption{ \textbf{Corner states excited by an in-plane AC magnetic field.}
Left: Classical ground-state magnetic texture of a $30\times 30$ spin lattice at $g\mu_\textrm{B} B_{z}/(JS)=0.3$. 
Right: Steady-state amplitude of oscillations of the spins under an additional in-plane magnetic field at resonance with the magnonic corner states. Spins near the corners are predominantly excited.}
\label{fig8}
\end{figure}

In this section, using the LLG equation, we show that the topologically protected magnonic corner states can be excited by AC magnetic fields. The initial state is taken as the ground-state magnetic texture of a $30\times 30$ spin lattice with open boundary conditions at $g\mu_\textrm{B}B_z/(JS)=0.3$. To excite a particular magnonic state, the frequency of the AC magnetic field should be at resonance with the energy of such state. From the diagonalization of the spin wave Hamiltonian, the energies of the corner states are found to be approximately at $\CalE/(JS) \approx 0.3$. Thus, we apply the additional in-plane magnetic field $\boldsymbol{B}_{\parallel}(t)=B_0\cos(\omega t)\xhat$ with frequency $\hbar\omega/(JS)=0.3$ and amplitude $g\mu_\textrm{B} B_0/(JS)=0.01$. In Fig.~\ref{fig8} we show the amplitude of oscillations of spins at each site, defined as
\begin{align}
|\delta \Bm_{\Br}| = \left[ \sum_{i=1}^3 ( \textrm{max}_{t} \{ m_{\Br}^{i}(t) \} - \textrm{min}_t \{ m_{\Br}^{i}(t) \} )^2 \right]^{1/2} \,,
\end{align}
with max$_t$ and min$_t$ evaluated in the interval $[t_0,t_0+T]$, i.e., over one period $T$ of the AC magnetic field after a long time $t_0=10^4$, to ensure that the system has reached a steady state. Although off-resonance bulk modes are weakly excited, Figure~\ref{fig8} clearly shows large oscillation amplitudes at the corners of the sample. We note that an out-of-plane AC magnetic field can also excite the magnonic corner states.


\section{Robustness of magnonic corner states against disorder}
\label{Appendix: disorder}

\begin{figure}[t]
\centering
\includegraphics[width=0.8\columnwidth]{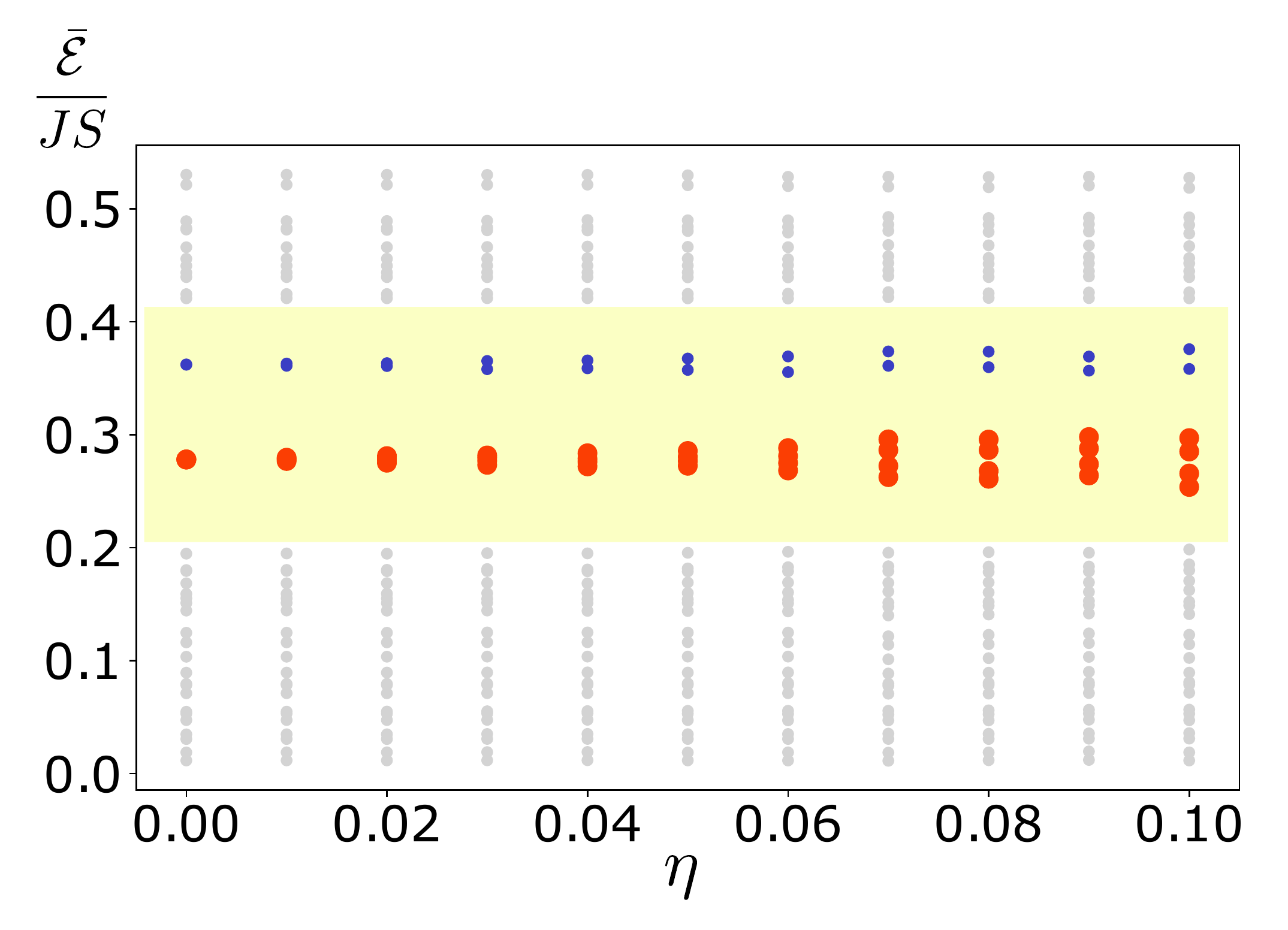}
\caption{ \textbf{Robustness of corner states against disorder.}
Disorder-averaged magnon spectrum, over 20 realizations, of a $30\times 30$ spin lattice as a function of the disorder strength $\eta$ at $g\mu_\textrm{B} B_{z}/(JS)=0.3$. The fourth bulk magnon gap is indicated in yellow while the energies of magnonic corner and edge-localized states are highlighted in red and blue, respectively.
}
\label{fig9}
\end{figure}

In order to confirm the expected robustness of the topologically protected magnonic corner states, we introduce disorder as random fluctuations in the $z$-component of the external magnetic field across the sample: $\delta B_z(\Br)$. These random fluctuations are added on top of the uniform, external magnetic field $B_z$. Taking into account the effect of a given disorder realization is a two-step process. First, the new classical ground-state magnetic texture is computed using Monte Carlo simulated annealing, which tends to take more Monte Carlo steps to be reached. Second, the disorder magnon spectrum is computed. This means that two modifications go into the construction of the spin wave Hamiltonian: the disordered texture must be used and the following term has to be added
\begin{align}
H_{\rm{dis}} = \sum_{\Br} g\mu_\textrm{B}\delta B_z(\Br) \zhat\cdot\Bm_{\Br} a^\dagger_{\Br} a_{\Br} \,.
\end{align}
We model the random fluctuations as $\delta B_z(\Br) = B_z\chi_{\Br}\eta$, where $\chi_{\Br}$ is uniformly distributed in the interval $[-1,1]$ and $\eta$ is a parameter that controls the disorder strength. In our study, we set the uniform, external magnet field at $g\mu_\textrm{B} B_{z}/(JS)=0.3$ so that fractional antiskyrmions are stable at all edges. Figure~\ref{fig9} shows the disorder-averaged magnon spectrum $\bar{\mathcal{E}}$ against the disorder strength $\eta$. For each value of $\eta$, we take a statistical average over $20$ disorder realizations. Although disorder breaks the fourfold degeneracy of the corner states, their disorder-averaged energies remain within the bulk magnon gap. Hence, we conclude that the magnonic corner states are robust against moderate disorder.


\section{Multipole moments in magnonic systems}

The quantum theory of multipole moments in magnonic systems can be constructed analogously to its electronic counterpart. However, a subtle yet critical modification must be implemented to account for the fact that the bosonic spin wave Hamiltonian is diagonalized by a paraunitary rather than a unitary matrix. We should also note that due to the Bogoliubov-de Gennes form of the spin wave Hamiltonian, the full magnon spectrum contains redundant, magnon ``hole'' bands \cite{shindou_topological_2013}. In practice, only the energy eigenstates corresponding to ``particle'' magnon bands are considered. In what follows, the lattice constant of the spin lattice is set to one for simplicity.


\subsection{Wilson loop and bulk polarization}
\label{Appendix: wilson loop}

We define the wave function of the $n$-th magnon band as $\ket{u_{\Bk}^n}=T_{\Bk}v_n$, where $T_{\Bk}$ is the paraunitary matrix that diagonalizes the spin wave Hamiltonian in reciprocal space, and $v_n$ is a vector whose components are given by $v_n^j = \delta_{nj}$. Since $T_{\Bk}$ is paraunitary, the orthogonality relation of magnonic wave functions is obtained by introducing the following inner product
\begin{align}
\braket{u_{\Bk}^m|u_{\Bk}^n}_\textrm{para}\equiv \xi v_m^T T^\dagger_{\Bk}\sigma_3  T_{\Bk}v_n=\delta_{mn} \,,
\label{dot_para}
\end{align}
with $\xi = \pm 1$ for particle/hole bands and $\sigma_3$ as in \eqref{eq:sigma3}. 

Using this orthogonal basis, we can introduce the Wilson loop $W_{x,\Bk}$ of the lowest $M$ bands in two dimensional systems as
\begin{align}
W_{x,(k_x,k_y)}&=F_{(k_x+(N_k-1)\Delta_{k},k_y)}\nonumber\\
&\times F_{(k_x+(N_k-2)\Delta_{k},k_y)}\hdots F_{(k_x,k_y)} \,,
\end{align}
where $\Delta_{k}=2\pi/N_k$ and
\begin{align}
[F_{(k_x,k_y)}]^{mn}=\braket{u_{(\Delta_{k}+k_x,k_y)}^m|u_{(k_x,k_y)}^n}_\textrm{para} \,,
\label{loop_element}
\end{align}
with $m,n=1,\hdots, M$. Here, we take the periodic gauge across the Brillouin zone. In this construction, as discussed below, the Wilson loop becomes unitary in the thermodynamic limit. In practice, for finite $N_k$, we can use the singular value decomposition of the $F_{\Bk}$ matrices to obtain a unitary Wilson loop. At any rate, the eigenvalues of the Wilson loop are ensured to be phase factors. In fact
\begin{align}
W_{x,\Bk}\ket{\nu_{x,\Bk}^j}=e^{i2\pi \nu_x^j(k_y)}\ket{\nu_{x,\Bk}^j} \,,
\label{wilson_loop}
\end{align}
where $j=1,\hdots, M$ and the $\nu_x^j(k_y)$ are the Wannier centers. We should note that the eigenstates $\ket{\nu_{x,\Bk}^j}$ depend on both $k_x$ and $k_y$ although the Wannier centers $\nu^j_x(k_y)$ only depend on $k_y$.

The hybrid Wannier functions along the $x$-axis are then given by
\begin{align}
\ket{\Psi^j_{R_x,k_y}}=\frac{1}{\sqrt{N_k}}\sum_{n=1}^{M}\sum_{k_x}[\nu_{x,\Bk}^j ]^n e^{-ik_xR_x}T_{\Bk}v_n \,,
\end{align}
where $R_x\in \{ 0,1,2, \hdots\}$ labels the unit cells along $x$-axis and $[\nu_{x,\Bk}^j ]^n$ is the $n$-th element of $\ket{\nu_{x,\Bk}^j}$. Since the Wannier functions are the Fourier transform of the Bloch functions, they form a complete orthogonal set, i.e.,
\begin{align}
\braket{\Psi^j_{R_x,k_y}|\Psi^{j'}_{R'_x,k'_y}}_\textrm{para}=\delta_{R_x,R'_x}\delta_{k_y,k'_y}\delta_{j,j'} \,.
\end{align} 
The Wannier center $\nu_x^j(k_y)$ is the expectation value of the position operator $\hat{x}$ evaluated in the Wannier functions~\cite{vanderbilt_2018}
\begin{align}
\nu_x^j(k_y)=\bra{\Psi^j_{0,k_y}}\hat{x}\ket{\Psi^{j}_{0,k_y}}_\textrm{para}  \textrm{ mod 1} \,.
\label{wannier_center}
\end{align}
In order to ensure that the Wannier centers obtained using Eq.~\eqref{wilson_loop} and Eq.~\eqref{wannier_center} agree, the origin for the eigenvalues of $\hat{x}$ must coincide with the origin defined via the Wilson loop. The latter corresponds to the site within the magnetic unit cell whose associated energy eigenstate component is purely real for all eigenstates. 

The bulk polarization is given by
\begin{align}
p_x(k_y) = \sum_{j=1}^M  \nu_x^j(k_y)  \textrm{ mod 1} \,,
\end{align}
and \emph{mutatis mutandis}
\begin{align}
p_y(k_x) = \sum_{j=1}^M  \nu_y^j(k_x)  \textrm{ mod 1} \,.
\end{align}
In electronic systems, the bulk polarization is proportional to the Berry phase in the thermodynamic limit. The same relationship also holds in magnonic systems. In the thermodynamic limit, we have
\begin{align}
[F_{\Bk}]^{mn}&=\xi v_m^T(T_{\Bk}^\dagger+\Delta_k \partial_{k_x} T_{\Bk}^\dagger)\sigma_3 T_{\Bk}v_n\nonumber\\
&=\delta_{mn}-\Delta_k \xi v_m^T T_{\Bk}^\dagger \sigma_3 \partial_{k_x} T_{\Bk}v_n \nonumber\\
&=\delta_{mn}+i\Delta_kA_{\Bk,x}^{mn} \,,
\end{align}
where $A_{\Bk,x}^{mn}$ is the non-Abelian Berry connection given by
\begin{align}
A_{\Bk,x}^{mn}=i\braket{u_{\Bk}^m|\partial_{k_x}|u_{\Bk}^n}_\textrm{para}=i\xi v_m^T T_{\Bk}^\dagger \sigma_3 \partial_{k_x} T_{\Bk}v_n \,.
\end{align}
We should note that this is a generalization of the Berry connection for a single band~\cite{shindou_topological_2013}. Hence, the Wilson loop is derived as
\begin{align}
W_{x,\Bk}&=\lim_{N_k\rightarrow\infty}\prod_{\ell=0}^{N_k-1} \left[I+i\Delta_k A_{(k_x+\ell\Delta_k,k_y),x}\right]\nonumber\\
&=\exp \left[i\int^{2\pi}_{0} A_{\Bk,x} \,dk_x \right] \,.
\end{align} 
Since the Berry connection is purely real, the Wilson loop matrix is unitary in the thermodynamic limit. Finally, the bulk polarization can be written as
\begin{align}
p_x(k_y) = \frac{1}{2\pi}\int^{2\pi}_{0}\textrm{tr}[A_{\Bk,x}]\,dk_x \,.
\end{align}


\subsection{Nested Wilson loop and bulk quadrupole moment}
\label{Appendix: nested}

The nested Wilson loop is defined in a similar manner. Given the gapped Wannier spectrum $\nu_{x}(k_y)$, we define a subspace of Wannier bands as $\nu_{x}^{+}\in [0,\frac{1}{2})$ and $\nu_x^-\in[-\frac{1}{2},0)$. Let us introduce the Wannier band basis
\begin{align}
\ket{\omega_{x,\Bk}^{j}}=\sum_{n=1}^{M}[\nu_{x,\Bk}^j]^nT_{\Bk}v_n \,,
\end{align}
where $j=1,\hdots, M_W$ with $M_W$ is the number of bands that belong to the Wannier sector $\nu_x^{\pm}$.
As in the previous section, the nested Wilson loop $\tilde{W}^{\nu_x^\pm}_{y,\Bk}$ is defined as
\begin{align}
[\tilde{W}^{\nu_x^{\pm}}_{y,\Bk}]^{j,j'}&=\braket{\omega_{x,(k_x,k_y+2\pi)}^j|\omega_{x,(k_x,k_y+(N_k-1)\Delta_k)}^r}_\textrm{para}\nonumber\\
&\times \bra{\omega_{x,(k_x,k_y+(N_k-1)\Delta_k)}^r} \hdots \ket{\omega_{x,(k_x,k_y+\Delta_k)}^t}_\textrm{para}\nonumber\\
&\times \braket{\omega_{x,(k_x,k_y+\Delta_k)}^t|\omega_{x,(k_x,k_y)}^{j'}}_\textrm{para} \,,
\end{align}
where the paraunitary matrix is inserted as defined in Eq.~(\ref{dot_para}) to ensure the orthogonality relations in the Wannier band basis. Diagonalizing the nested Wilson loop, we obtain
\begin{align}
\tilde{W}^{\nu_x^{\pm}}_{y,\Bk}\ket{\nu_{y,\Bk}^{\nu_x^{\pm}, p}}=e^{i2\pi \nu_y^{\nu_x^{\pm},p}(k_x)}\ket{\nu_{y,\Bk}^{\nu_x^{\pm}, p}} \,,
\label{nested_center}
\end{align}
where $\nu_{y}^{\nu_x^{\pm},p}$ is the Wannier center along $y$-axis evaluated in the Wannier sector $\nu_x^{\pm}$ and $p\in 1... M_W$. Alternatively, the eigenvalues of the nested Wilson loop can be written as
\begin{align}
\nu_{y}^{\nu_x^{\pm},p}(k_x)=\bra{\Phi^{\nu_x^{\pm},p}_{0,k_x}}\hat{y}\ket{\Phi^{\nu_x^{\pm},p}_{0,k_x}}_\textrm{para} \textrm{ mod 1} \,,
\end{align}
with
\begin{align}
\ket{\Phi^{\nu_x^{\pm},p}_{R_y,k_x}}=\frac{1}{\sqrt{N_k}}\sum_{k_y}\sum_{j=1}^{M_W}[\nu_{y,\textbf{k}}^{\nu_x^{\pm}, p}]^j e^{-ik_yR_y}\ket{\omega_{x,(k_x,k_y)}^{j}} \,,
\end{align}
where $R_y\in \{0,1,2,\ldots\}$ labels the unit cells along the $y$-axis and $[\nu_{y,\textbf{k}}^{\nu_x^{\pm}, p}]^j $ is the $j$-th element of $\ket{\nu_{y,\Bk}^{\nu_x^{\pm}, p}}$. The total Wannier sector polarization is
\begin{align}
p_y^{\nu_x^\pm}=\frac{1}{N_k}\sum_{k_x}\sum_{p=1}^{M_W}\nu_{y}^{\nu_x^\pm,p}(k_x) \textrm{ mod 1} \,,
\label{nested_pol}
\end{align}
and similarly
\begin{align}
p_x^{\nu_y^\pm}=\frac{1}{N_k}\sum_{k_y}\sum_{p=1}^{M_W}\nu_{x}^{\nu_y^\pm,p}(k_y) \textrm{ mod 1} \,.
\label{nested_pol}
\end{align}

Finally, the bulk quadrupole moment is constructed from the Wannier sector polarizations and is given by~\cite{benalcazar_electric_2017}
\begin{align}
q_{xy}=p_y^{\nu_x^+}p_x^{\nu_y^+}+p_y^{\nu_x^-}p_x^{\nu_y^-} \,.
\end{align}


\section{Higher-order topology in antiskyrmion crystals}


\subsection{Symmetry constraints and quantized bulk quadrupole moment}
\label{Appendix: symmetry}

In this section, we discuss the symmetries of the antiskyrmion crystal that quantize the bulk quadrupole moment and protect the magnonic corner states. The magnetic unit cell of the antiskyrmion crystal is left invariant under the following operations: $C_{2x}\mathcal{T}$, twofold rotation about the $x$-axis together with time reversal; $C_{2y}\mathcal{T}$, twofold rotation about the $y$-axis together with time reversal; and $C_{2z}$, twofold rotation about the $z$-axis [see Fig.~\ref{fig3}(a)].

The $C_{2y}\mathcal{T}$ operation transforms the spin wave Hamiltonian $H_{\Bk}$ as
\begin{align}
g_{\Bk}KH_{(k_x,k_y)}Kg_{\Bk}^\dagger=H_{(k_x,-k_y)} \,,
\end{align}
where $K$ is the complex conjugation operator and $g_{\Bk}$ is the unitary operator representation of $C_{2y}$.
This leads to the symmetry protected degeneracy for $k_y=\pi/L_ya$. Similarly, the double degeneracy is protected by $C_{2x}\mathcal{T}$ at $k_x=\pi/L_xa$. Here, $L_x$ and $L_y$ are the number of lattice sites within the antiskyrmion crystal magnetic unit cell along the $x$- and $y$-axis, respectively. This explains the double degeneracy from $X$ to $M$ and  $X'$ to $M$ in Fig.~\ref{fig3}(b). 

Following Ref.~\cite{benalcazar_electric_2017}, we derive the effect of the symmetry operation $C_{2y}\mathcal{T}$ on the Wilson loop. Let us introduce the unitary sewing matrix that connects two degenerate states at $\Bk$ transformed by $C_{2y}\mathcal{T}$
\begin{align}
B_{C_{2y}\mathcal{T},\Bk}^{mn}&=\bra{u_{(k_x,-k_y)}^m}g_{\Bk}K \ket{u_{(k_x,k_y)}^n}_\textrm{para}\nonumber\\
&=\xi v_m^T T_{(k_x,-k_y)}^{\dagger}g_{\Bk}K \sigma_3T_{(k_x,k_y)}v_n \,.
\end{align}
From $g_{\Bk}K\ket{u_{(k_x,k_y)}^n}=\ket{u_{(k_x,-k_y)}^m} B_{C_{2y}\mathcal{T},\Bk}^{mn}$, we have
\begin{align}
\ket{u_{(k_x,k_y)}^{n*}}=g_k^\dagger \ket{u_{(k_x,-k_y)}^m}B_{C_{2y}\mathcal{T},\Bk}^{mn} \,.
\end{align}
Taking the complex conjugate of both sides and noting that $[\sigma_3,g_{\Bk}]=[\sigma_3,K]=0$, we obtain the symmetry transformation for a Wilson loop element
\begin{align}
F_{(k_x,k_y)}^{mn}&=\braket{u_{(k_x+\Delta_k,k_y)}^m|u_{(k_x,k_y)}^n}_\textrm{para}\nonumber\\
&=[B_{C_{2y}\mathcal{T},(k_x+\Delta_k,k_y)}^T\nonumber\\
&\times \braket{u_{(k_x+\Delta_k,-k_y)}^{*}|g_k^*g_k^T |u_{(k_x,-k_y)}^*}_\textrm{para}\nonumber\\
&\times B_{C_{2y}\mathcal{T},(k_x,k_y)}^*]^{mn} \,.
\end{align}
The symmetry transformation of the Wilson loop is then given by
\begin{align}
W_{x,(k_x,k_y)}=B_{C_{2y}\mathcal{T},(k_x,k_y)}^T W_{x,(k_x,-k_y)}^* B_{C_{2y}\mathcal{T},(k_x,k_y)}^* \,.
\end{align}
Given $W_{x,\Bk}\ket{\nu_{\Bk}^j}=e^{i2\pi \nu_x^j(k_y)}\ket{\nu_{\Bk}^j}$, we obtain
\begin{align}
W_{x,\Bk}B_{C_{2y}\mathcal{T},\Bk}^T\ket{\nu_{\Bk}^j}&=B_{C_{2y}\mathcal{T},\Bk}^T W_{x,(k_x,-k_y)}^* \ket{\nu_{\Bk}^j}\nonumber\\
&=e^{-i2\pi\nu_x^j(-k_y)}B_{C_{2y}\mathcal{T},\Bk}^T\ket{\nu_{\Bk}^j} \,.
\end{align}
This implies that
\begin{align}
\nu_x^j(k_y)\overset{C_{2y}\mathcal{T}}{=}-\nu_x^j(-k_y) \textrm{ mod }1 \,.
\end{align}
Likewise, the Wannier center in the $y$-axis is constrained by 
\begin{align}
\nu_y^j(k_x)\overset{C_{2x}\mathcal{T}}{=}-\nu_y^j(-k_x)\textrm{ mod }1 \,.
\end{align}
Therefore, the in-plane rotation symmetry combined with time reversal symmetry leads to $\nu_x=0, \frac{1}{2}$ or a pair of eigenvalues satisfying $\nu_x^1(k_y)=-\nu_x^2(-k_y)$. We should note that the $C_{2z}$ symmetry also gives the same constraint on the Wannier center~\cite{benalcazar_electric_2017}. After summing over the Wannier centers, the polarization $p_x$ and $p_y$ are quantized to be either $0$ or $\frac{1}{2}$.

Similarly, the symmetry constraints on the eigenvalues of the nested Wilson loop are derived as
\begin{align}
\nu_x^{\nu_y, j}(k_y)&\overset{C_{2y}\mathcal{T}}{=}-\nu_x^{\nu_y, j}(-k_y) \textrm{ mod }1 \,, \label{s1} \\
\nu_y^{\nu_x,j}(k_x)&\overset{C_{2x}\mathcal{T}}{=}-\nu_y^{\nu_x,j}(-k_x)\textrm{ mod }1 \,. \label{s2}
\end{align}
Hence, the Wannier sector polarization is quantized to be 0 or $\frac{1}{2}$.
In addition, the $C_{2z}$ symmetry relates the Wannier centers from different Wannier sectors via
\begin{align}
\nu_x^{\nu_y, j}(k_y)&\overset{C_{2z}}{=}-\nu_x^{-\nu_y, j}(-k_y) \textrm{ mod }1 \,, \label{s3}\\
\nu_y^{\nu_x, j}(k_x)&\overset{C_{2z}}{=}-\nu_y^{-\nu_x, j}(-k_x) \textrm{ mod }1 \,. \label{s4}
\end{align}
From the above two Eqs. is follows that the Wannier sector polarizations satisfy
\begin{align}
p_x^{\nu_y}&=-p_x^{-\nu_y} \textrm{ mod }1 \,, \\
p_y^{\nu_x}&=-p_y^{-\nu_x} \textrm{ mod }1 \,.
\end{align}
Finally, the bulk quadrupole moment is given by
\begin{align}
q_{xy}&=p_y^{\nu_x^+}p_x^{\nu_y^+}+p_y^{\nu_x^-}p_x^{\nu_y^-}=2p_y^{\nu_x^-}p_x^{\nu_y^-}\nonumber\\
&=0\textrm{ or } \frac{1}{2} \textrm{ mod }1 \,.
\end{align}
As shown in the Fig.~\ref{fig3}, the Wannier sector polarizations of the lowest four bands magnon bands of the antiskyrmion crystal are obtained as $p_y^{\nu_x^-}=p_x^{\nu_y^-}=-\frac{1}{2} \textrm{ mod }1$. Therefore, the bulk quadrupole moment of the antiskyrmion crystal is quantized to $q_{xy} = \frac{1}{2}$.


\subsection{Edge dipole moment in a strip geometry}
\label{Appendix:Edge}

\begin{figure}[t]
\centering
\includegraphics[width=0.8\columnwidth]{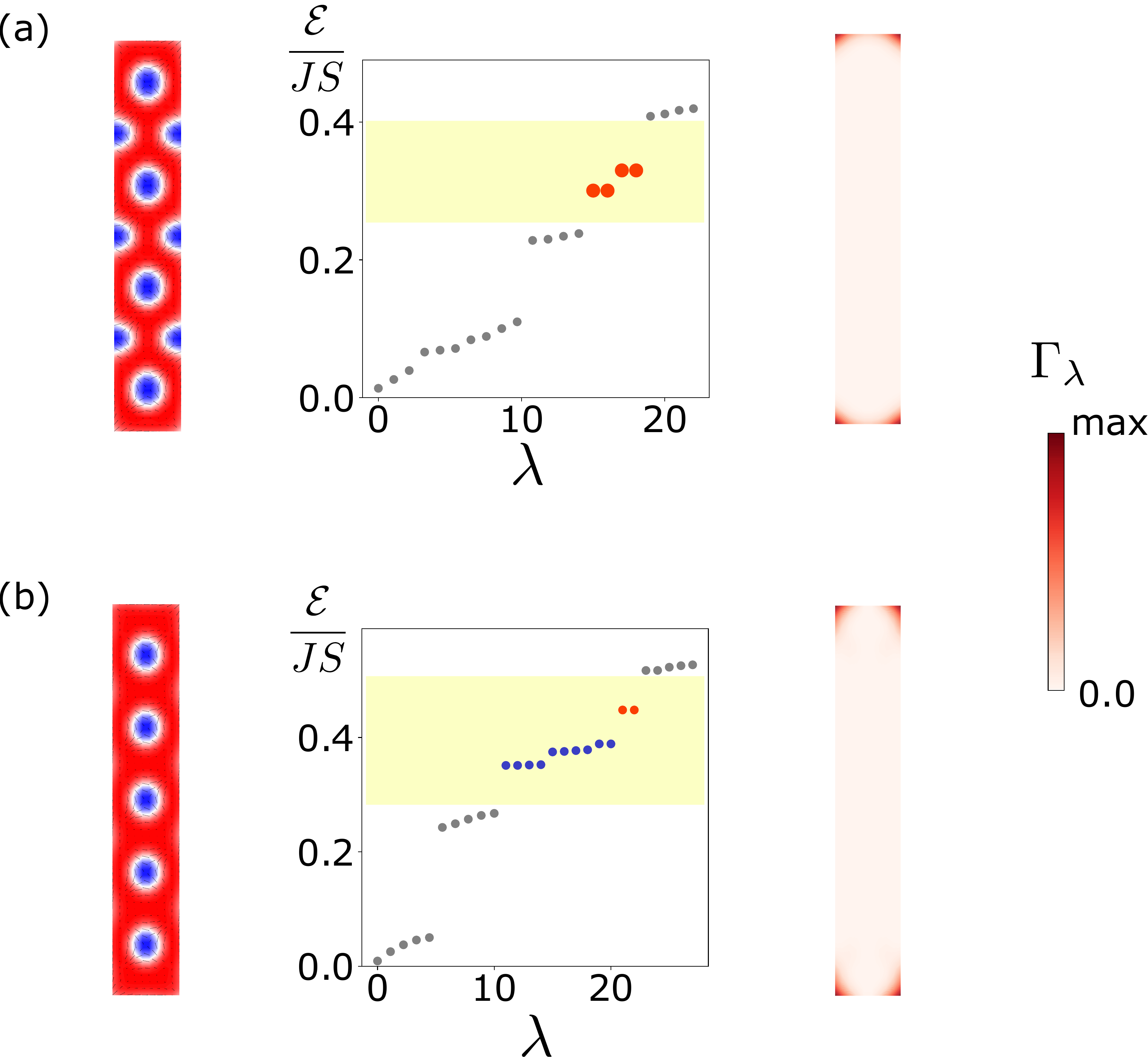}
\caption{ \textbf{Edge polarization induces corner states below the edge instability in finite systems.} The parameters are the same as in Fig.~\ref{fig4} with (a) $g\mu_\textrm{B}B_z/(JS)=0.3$ and (b) $g\mu_\textrm{B}B_z/(JS)=0.41$. Left: classical ground-state spin textures. Middle: magnon spectrum showing corner states (red) and edge states (blue). Right: probability density of the lowest energy corner state. }
\label{fig10}
\end{figure}

Let us consider a system that is periodic along the $x$-axis while of finite extension and with open boundary conditions along the $y$-axis. Similarly to the calculation of the bulk polarization, we construct the Wilson loop along the $x$-axis as
\begin{align}
W_{x,k_x}=F_{k_x+(N_k-1)\Delta_{k}}F_{k_x+(N_k-2)\Delta_{k}}\hdots F_{k_x} \,,
\end{align}
where $\Delta_{k}=2\pi/(L_xaN_k)$ and
\begin{align}
[F_{k_x}]^{mn}=\braket{u_{\Delta_{k}+k_x}^m|u_{k_x}^n}_\textrm{para} \,,
\label{loop_element}
\end{align}
with $m,n=1,\hdots, M'$. Here, the number of occupied bands is typically given by $M'=M\times N_y$ where $N_y$ is the number unit cells in the $y$-axis. However, we should not include topologically trivial edge bands that may appear above the edge instability as they are not bulk bands. After diagonalizing the Wilson loop, we have
\begin{align}
W_{x,k_x}\ket{\nu_{x,k_x}^j}=e^{i2\pi \nu_x^j}\ket{\nu_{x,k_x}^j} \,,
\end{align}
where $j=1,\hdots M'$ and $\nu_x^j$ are the Wannier centers. The hybrid Wannier functions along the $x$-axis are then given by
\begin{align}
\ket{\Psi^j_{R_x}}=\frac{1}{\sqrt{N_k}}\sum_{n=1}^{M'}\sum_{k_x}[\nu_{x,k_x}^j ]^n e^{-ik_xR_x}T_{k_x}v_n \,,
\end{align}
where $R_x\in \{0,1,2,\ldots \}$ labels the unit cells along the $x$-axis and $[\nu_{x,k_x}^j ]^n$ is the $n$-th element of $\ket{\nu_{x,k_x}^j}$. The polarization within each unit cell is given by \cite{benalcazar_electric_2017}
\begin{align}
p_x(R_y)=\sum_{j=1}^{M'}\nu_x^j \,\rho^j(R_y) \,,
\end{align}
where $\rho^j(R_y)$ is the probability density of the hybrid Wannier function within the unit cell labeled by $R_y \in \{0,1,\hdots, N_y-1\}$.

\begin{figure}[t]
\centering
\includegraphics[width=0.95\columnwidth]{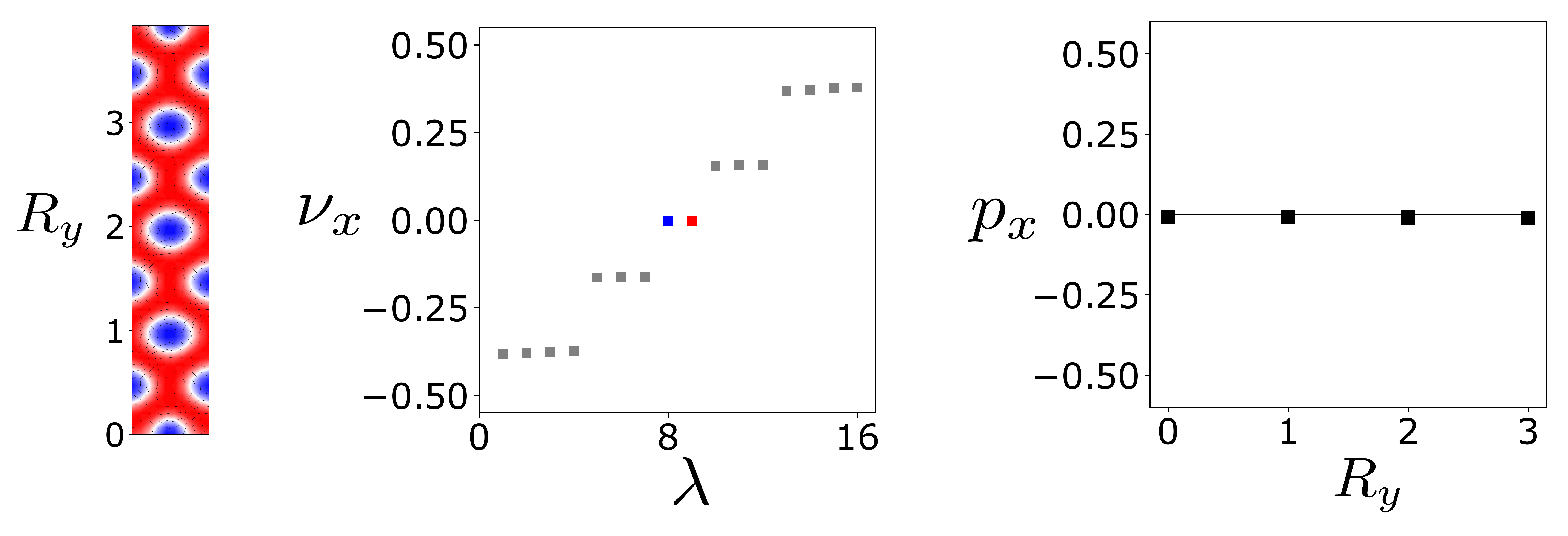}
\caption{ \textbf{Magnetic unit cell dependence of edge polarization.} Left: same magnetic unit cell as in Fig.~\ref{fig4}(a) but shifted horizontally by half its length. Middle: Wannier spectrum $\nu_x$ showing four bulk bands (grey) and the edge modes (red and blue) which are at zero. Right: the polarization $p_x(R_y)$ averaged over each vertical unit cell vanishes. 
 }
\label{fig11}
\end{figure}

As shown in Fig.~\ref{fig4}, quantized edge dipole moments are obtained below the edge instability where, due to the fractional antiskyrmions stabilized along the strip edges, the symmetries of the the antiskyrmion crystal are preserved. In contrast, the edge dipole moment vanishes above the edge instability as fractional antiskyrmions that support Wannier edge states are no longer stable. The fact that a quantized edge dipole translates into magnonic corner charges can be confirmed by turning the strip geometry unit cell into a fully finite system with open boundary conditions as depicted in Fig.~\ref{fig10}(a). We should note that Fig.~\ref{fig10}(b) shows trivial corner states that are twofold degenerate and merge with bulk bands at higher magnetic fields.

The edge polarization also depends on the choice of unit cell \cite{watanabe_inequivalent_2018}. As shown in Eq.~(\ref{wannier_center}), the Wannier center is the expectation value of the position operator. If we shift the choice of unit cell by half a magnetic unit cell, the Wannier centers are shifted by $\frac{1}{2}$. This is demonstrated in Fig.~\ref{fig11}, showing the Wannier spectrum and polarization of a strip magnetic unit cell shifted by half a unit cell with respect to the one shown in Fig.~\ref{fig4}(a). In this new configuration, the edge polarization vanishes because of the shift of the Wannier centers, which localize at the cores of the edge fractional antiskyrmions, from $\frac{1}{2}$ to $0$. As a result, it would not host magnonic corner states if boundaries orthogonal to the $x$-axis were introduced. We should note that this choice of magnetic unit cell is energetically less favorable so the bulk-boundary correspondence holds for the classical ground state in the thermodynamic limit. However, a local configuration equivalent to Fig~\ref{fig11} can be realized due to finite size effects as shown in Fig.~\ref{fig5}~(c).


\subsection{Corner charges induced by the bulk quadrupole moment}
\label{Appendix: Qc}

Although magnons carry no electric charge, we can still introduce a quantity similar to the electric corner charges constructed out of magnon number densities. We draw inspiration from the concept of boundary charge studied in one-dimensional systems, which is defined as the difference between the charge near the boundary and the background charge from the bulk states~\cite{suSolitonsPolyacetylene1979a, suFractionallyChargedExcitations1981a, jackiwSolitonsFermionNumber1981, park_fractional_2016, thakurathi_fractional_2018, pletyukhov_topological_2020, pletyukhov_surface_2020}. 
  
Let us consider a finite system that contains $N_x\times N_y$ magnetic unit cells. The magnon charge density carried by the lowest $M$ bulk magnon bands of the antiskyrmion crystal is defined as
\begin{align}
\bar{\rho}(\Br)&= \sum_{n=1}^{N_x\times N_y\times M}  \braket{u_{\Br}^n|u_{\Br}^n}_\textrm{para}  - \braket{\tilde{u}_{\Br}^n|\tilde{u}_{\Br}^n}_\textrm{para} \,,
\label{eq:rho}
\end{align}
where $\braket{u_{\Br}^n|u_{\Br}^n}_\textrm{para}$ is the magnon density of the $n$-th magnonic wave function at site $\Br$ on this finite system. As a convenient bulk-like magnon density we use $\braket{\tilde{u}_{\Br}^n|\tilde{u}_{\Br}^n}_\textrm{para}$, where $\ket{\tilde{u}_{\Br}^n}$ is the $n$-th magnonic wave function, evaluated at $k_x = 0$, of an auxiliary periodic system, obtained from the finite system here considered by connecting the edges orthogonal to the $x$-axis. From the magnon charge density we can construct the magnon charge at each magnetic unit cell as
\begin{align}
Q_{(R_x,R_y)}=\sum_{\Br\in (R_x,R_y)}\bar{\rho}(\Br) \,,
\label{eq:Q_c}
\end{align}
where $R_x=0,\hdots, N_x-1$ and $R_y=0,\hdots, N_y-1$ label the magnetic unit cells. 

\begin{figure}[t]
\centering
\includegraphics[width=\columnwidth]{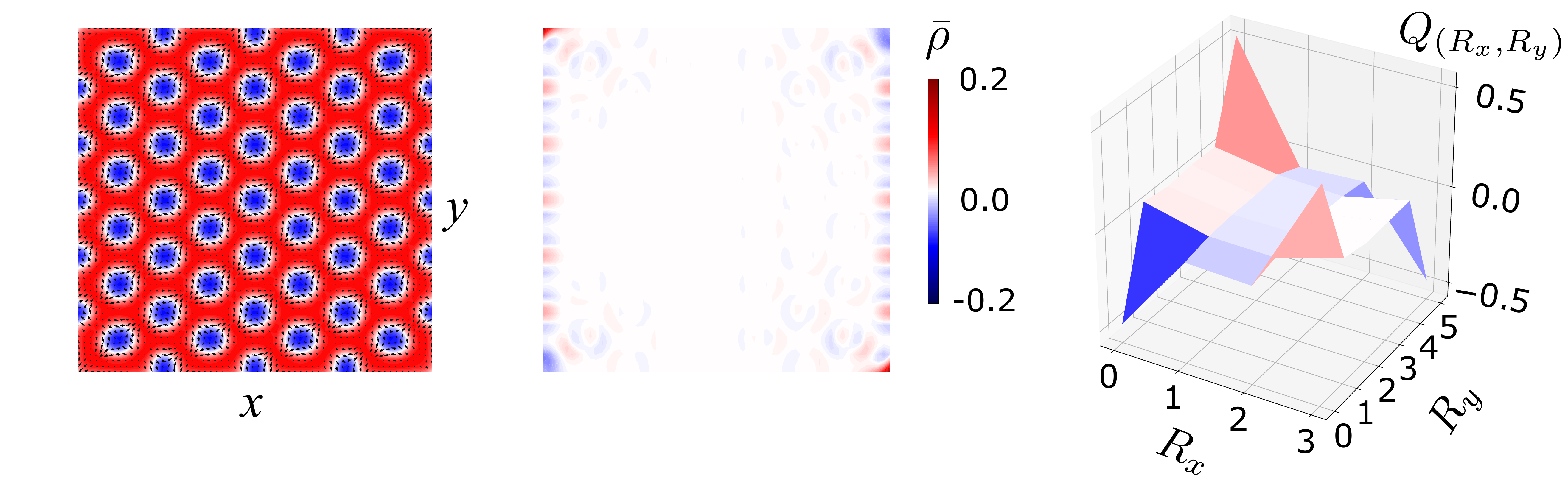}
\caption{ \textbf{Corner charges induced by the bulk quadrupole moment.} Left: classical ground-state magnetic texture of a $60\times 60$ spin lattice at $g\mu_\textrm{B}B_z/(JS)=0.3$. Middle: magnon charge density $\bar{\rho}$. The lowest $N_x\times N_y\times M$ states are occupied with $N_x\times N_y=24$ and $M=4$. Right: magnon charge $Q_{(R_x,R_y)}$. The color indicates positive (red) and negative (blue) charges.}
\label{fig12}
\end{figure}

We compute the magnon corner of a $60\times 60$ spin lattice with open boundary conditions at $g\mu_\textrm{B}B_z/(JS)=0.3$. As shown in the left panel of Fig.~\ref{fig12}, it contains $24$ bulk magnetic unit cells with $N_x=4$ and $N_y=6$. Since the quantized bulk quadrupole moment of the antiskyrmion crystal involves the lowest four bulk magnon bands, we take $M=4$. The corresponding auxiliary, periodic system is prepared by the identification procedure described above and then letting the magnetic texture relax using the LLG equation for a long time $t_0=10^4$. In order to unambiguously define the magnon corner charge, the fourfold degeneracy of the magnonic corner states must be slightly broken. This is achieved by introducing a single disorder realization with disorder strength $\eta = 0.1$, as described in Appendix \ref{Appendix: disorder}. Also shown in Fig.~\ref{fig12} are the magnon charge density $\bar{\rho}$ (middle) and the magnon charge $Q_{(R_x,R_y)}$ (right). In the middle panel, we observe boundary charges along the vertical edges as expected for the auxiliary, periodic system employed. As we move away from these edges, the magnon density of the finite system becomes identical to the bulk-like magnon density and the magnon charge density converges to zero. The right panel clearly shows that the magnon charge is corner-localized and quantized as $|Q_c|=\frac{1}{2}$ with diagonally opposite corners having the same sign of magnon charge.

\bibliography{MagQuadTIAntiSkXs}

\end{document}